\documentclass[11pt,reqno]{article}

\DeclareSymbolFont{eulerletters}{U}{eur}{m}{n}%
\DeclareMathSymbol{\PowersetSym}{\mathord}{eulerletters}{"7D}%
\newsavebox{\powersetbox}
\sbox{\powersetbox}{\mbox{\Large\ensuremath{\PowersetSym}}}
\providecommand{\powerset}{\mathopen{\usebox{\powersetbox}}}

\usepackage{ulem}
\usepackage{comment}

\usepackage[usenames,dvipsnames]{color}

\definecolor{pink}{rgb}{1.00,0.08,0.58}

\usepackage{epsfig} \usepackage{xspace}
\usepackage{color}
\usepackage{boxedminipage}

\usepackage{amsmath}
\usepackage{amssymb}

\usepackage{pifont}

\let\ensuremathTEMP\ensuremath

\makeatletter%
\def\nottoobig#1{{\hbox{$\left#1\vcenter
to1.111\ht\strutbox{}\right.\n@space$}}}
\makeatother%

\makeatother%

\makeatletter%

\newcount\hour \newcount\minutes \hour=\time \divide\hour by 60
\minutes=\hour \multiply\minutes by -60 \advance\minutes by \time
\def\mmmddyyyy{\ifcase\month\or Jan\or Feb\or Mar\or Apr\or May\or
Jun\or Jul\or Aug\or Sep\or Oct\or Nov\or Dec\fi \space\number\day,
\number\year}
\def\hhmm{\ifnum\hour<10 0\fi\number\hour :%
  \ifnum\minutes<10 0\fi\number\minutes} \def\Draft{{\it Draft of
\mmmddyyyy}}

\def\ps@jtsheadings{%
\def\@oddhead{\it\rightmark\hfil\rm\thepage}%
\def\@oddfoot{\hfil\Draft}%
\if@twoside%
\def\@evenhead{\rm\thepage\hfil\it\leftmark}%
\def\@evenfoot{\Draft\hfil}%
\else
\let\@evenhead\@oddhead%
\let\@evenfoot\@oddfoot%
\fi%
}
\def\ps@jtsplain{%
\def\@oddhead{\hfil\Draft}%
\def\@oddfoot{\hfil\rm\thepage\hfil}%
\let\@evenfoot\@oddfoot%
\if@twoside \def\@evenhead{\Draft\hfil} \else \let\@evenhead\@oddhead
\fi }

\def\chaptermark#1{\markboth{\thechapter.\ #1}{\thechapter.\ #1}}%
\def\sectionmark#1{\markright{\thesection.\ #1}}

\def\section{\@startsection {section}{1}{\z@} {3.5ex plus1ex
    minus.2ex}{2.3ex plus.2ex}{\Large\bf}}
    \def\subsection{\@startsection{subsection}{2}{\z@} {3.25ex plus1ex
    minus.2ex}{1.5ex plus.2ex}{\large\bf}}
    \def\subsubsection{\@startsection{subsubsection}{3}{\z@} {3.25ex
    plus1ex minus.2ex}{1.5ex plus.2ex}{\normalsize\bf}}
    \def\paragraph{\@startsection{paragraph}{4}{\z@} {3.25ex plus1ex
    minus.2ex}{1em}{\normalsize\bf}}
    \def\subparagraph{\@startsection{subparagraph}{4}{\parindent}
    {3.25ex plus1ex minus.2ex}{1em}{\normalsize\bf}}

\makeatother%

\makeatletter \@beginparpenalty=10000 \makeatother

\def\underl#1 {\leavevmode\let\first=\relax\underli #1 }
\def\underli#1 {\ifx&#1\let\next=\relax\unskip
\else\let\next=\underli\first\ulinebox{#1}\fi\let\first=\undersp\next}
\def\undersp{\penalty50\ulinebox{\space}\penalty50}
\def\ulinebox#1{\vtop{\hbox{\strut#1}\hrule}}%
\def\unice#1 {\underl #1 & }
\def\desclabel#1{\bf #1\hfil}
\def\desc{\list{}{%
\setlength{\leftmargin}{0pt} \labelwidth= \leftmargin \advance
\labelwidth by -\labelsep \let \makelabel=\desclabel}}

\def\descHACKlabel#1{\bf #1\hfil}
\def\descHACK{\list{}{%
\setlength{\leftmargin}{0pt} \labelwidth= \leftmargin \advance
\labelwidth by -\labelsep \let \makelabel=\descHACKlabel}}

\newcounter{extremeleftlistcounter}
  {\begin{list}{\arabic{extremeleftlistcounter}~~~}{\usecounter{extremeleftlistcounter}%
        \setlength{\labelsep}{0pt}\setlength{\leftmargin}{0pt}%
        \setlength{\labelwidth}{0pt}\setlength{\listparindent}{0pt}}}%
  {\end{list}}

\newcounter{leftlistcounter}
  {\begin{list}{\arabic{leftlistcounter}~~~}{\usecounter{leftlistcounter}%
        \setlength{\labelsep}{0pt}\setlength{\leftmargin}{15pt}%
        \setlength{\labelwidth}{15pt}\setlength{\listparindent}{0pt}}}%
  {\end{list}}



\makeatletter %

\newlength{\filength} 
\newsavebox{\gcbox}
\sbox{\gcbox}{\framebox[\filength]{\rule{0ex}{2ex}}}

\newlength{\leftjustindent} \newlength{\@leftjustindent}
\def\leftjust{\let\\\@leftjustcr\let\end\@endleftjust
\addtolength{\@leftjustindent}{\leftjustindent} \vcenter\bgroup
\halign\bgroup \hbox to\displaywidth{
\rule{\@leftjustindent}{0ex}$\displaystyle##$\hfill }\crcr }
\def\endleftjust{\crcr\egroup\egroup\endgroup}
\def\@endleftjust#1{\crcr\egroup\egroup\@checkend{#1}\endgroup}
\def\@leftjustcr{\crcr}

\newcommand{\red}[3]{ { {\rm R}_{#1}^{#2}({#3}) } }

\newtheorem{theorem}{Theorem}[section]

\newtheorem{corollary}[theorem]{Corollary}
\newtheorem{claim}{Claim}

\newcommand{\qedblob}{\mbox{\rule[-1.5pt]{5pt}{10.5pt}}}

\def\literalqed{{\ \nolinebreak\hfill\mbox{\qedblob\quad}}}
 \def\qed{\literalqed} 
\newtheorem{lemma}[theorem]{Lemma}

\newcommand{\singlespacing}{\let\CS=
\@currsize\renewcommand{\baselinestretch}{1}\tiny\CS}
\newcommand{\singlespacingplus}{\let\CS=
\@currsize\renewcommand{\baselinestretch}{1.25}\tiny\CS}
\newcommand{\doublespacing}{\let\CS=
\@currsize\renewcommand{\baselinestretch}{1.75}\tiny\CS}
\newcommand{\draftspacing}{\let\CS=
\@currsize\renewcommand{\baselinestretch}{2.0}\tiny\CS}
\newcommand{\hugedraftspacing}{\let\CS=
\@currsize\renewcommand{\baselinestretch}{2.4}\tiny\CS}
\newcommand{\normalspacing}{\singlespacing}
\makeatother%

\mathcode`\0="0030      %
\mathcode`\1="0031 \mathcode`\2="0032 \mathcode`\3="0033
\mathcode`\4="0034 \mathcode`\5="0035 \mathcode`\6="0036
\mathcode`\7="0037 \mathcode`\8="0038 \mathcode`\9="0039

\newtheorem{definition}[theorem]{Definition}

\makeatletter \clubpenalty=\@highpenalty \widowpenalty=\@highpenalty
\makeatother

\makeatletter
\newcommand{\niceonespacing}{\let\CS=\@currsize\renewcommand{\baselinestretch}{1.1}\tiny\CS}\newcommand{\nicetwospacing}{\let\CS=\@currsize\renewcommand{\baselinestretch}{1.2}\tiny\CS}
\newcommand{\nicethreespacing}{\let\CS=\@currsize\renewcommand{\baselinestretch}{1.3}\tiny\CS}
\newcommand{\singlespacingplusplus}{\let\CS=\@currsize\renewcommand{\baselinestretch}{1.35}\tiny\CS}
\newcommand{\nicefourspacing}{\let\CS=\@currsize\renewcommand{\baselinestretch}{1.4}\tiny\CS}
\newcommand{\nicefivespacing}{\let\CS=\@currsize\renewcommand{\baselinestretch}{1.5}\tiny\CS}
\newcommand{\nicesixspacing}{\let\CS=\@currsize\renewcommand{\baselinestretch}{1.6}\tiny\CS}
\makeatother

\makeatletter \def\@cite#1#2{[#1\if@tempswa , #2\fi]} \makeatother

\makeatletter
\def\@citex[#1]#2{\if@filesw\immediate\write\@auxout{\string\citation{#2}}\fi
\def\@citea{}\@cite{\@for\@citeb:=#2\do
{\@citea\def\@citea{,\linebreak[0]}\@ifundefined {b@\@citeb}{{\bf
?}\@warning
       {Citation `\@citeb' on page \thepage \space undefined}}%
\hbox{\csname b@\@citeb\endcsname}}}{#1}} \makeatother

\makeatletter
\def\ps@thesis{\def\@oddhead{\hfil\rm\thepage\hfil}\def\@oddfoot{}\def\@evenhead{\hfil\rm\thepage\hfil}\def\@evenfoot{}\def\chaptermark##1{}\def\sectionmark##1{}}
\makeatother

\makeatletter
\def\foobarpt{\textfont\z@\tenrm \scriptfont\z@\ninrm
  \scriptscriptfont\z@\sevrm \textfont\@ne\tenmi \scriptfont\@ne\ninmi
  \scriptscriptfont\@ne\sevmi \textfont\tw@\tensy
  \scriptfont\tw@\ninsy \scriptscriptfont\tw@\sevsy
  \textfont\thr@@\tenex \scriptfont\thr@@\tenex
  \scriptscriptfont\thr@@\tenex
  \def\unboldmath{\everymath{}\everydisplay{}\@nomath\unboldmath
  \textfont\@ne\tenmi \textfont\tw@\tensy \textfont\lyfam\tenly
  \@boldfalse}\@boldfalse
  \def\boldmath{\@ifundefined{tenmib}{\global\font\tenmib\@mbi\@magscale1\global
  \font\tensyb\@mbsy \@magscale1\global\font
  \tenlyb\@lasyb\@magscale1\relax\@addfontinfo\@xiipt
  {\def\boldmath{\everymath
  {\mit}\everydisplay{\mit}\@prtct\@nomathbold \textfont\@ne\tenmib
  \textfont\tw@\tensyb
                \textfont\lyfam\tenlyb\@prtct\@boldtrue}}}{}\@xiipt\boldmath}%
\def\prm{\fam\z@\tenrm}%
\def\pit{\fam\itfam\tenit}\textfont\itfam\tenit
   \scriptfont\itfam\ninit \scriptscriptfont\itfam\sevit
   \def\psl{\fam\slfam\tensl}\textfont\slfam\tensl
   \scriptfont\slfam\tensl \scriptscriptfont\slfam\tensl
   \def\pbf{\fam\bffam\tenbf}\textfont\bffam\tenbf
   \scriptfont\bffam\ninbf \scriptscriptfont\bffam\ninbf
   \def\ptt{\fam\ttfam\tentt}\textfont\ttfam\tentt
   \scriptfont\ttfam\nintt \scriptscriptfont\ttfam\nintt
   \def\psf{\fam\sffam\tensf}\textfont\sffam\tensf
   \scriptfont\sffam\tensf \scriptscriptfont\sffam\tensf
\def\psc{\@getfont\psc\scfam\@xiipt{\@mcsc\@magscale1}}%
\def\ly{\fam\lyfam\tenly}\textfont\lyfam\tenly \scriptfont\lyfam\ninly
   \scriptscriptfont\lyfam\sevly \@setstrut \rm}

\makeatother

\newcommand{\naturalnumber}{\ensuremath{{ \mathbb{N} }}}

\newcommand{\sharpp}{{\rm \#P}} 
 
 \newcommand{\up}{{\rm UP}}
 
\newcommand{\fewp}{{\rm FewP}} \newcommand{\coup}{{\rm coUP}}

 \newcommand{\nexp}{{\rm NEXP}}
\newcommand{\p}{{\rm P}} 
 \newcommand{\np}{{\rm NP}}

     \newcommand{\bpp}{{\rm BPP}}

\newcommand{\conp}{{\rm coNP}} \newcommand{\pspace}{{\rm PSPACE}}

\newcommand{\ph}{\ensuremath{{\rm PH}}}
\newcommand{\few}{{\ensuremath{\rm Few}}}

\newcommand{\sproof}{\noindent{\bf Proof}\quad}

\newcommand{\sigmastar}{\ensuremath{\Sigma^\ast}}

\newcommand{\condition}{\,\ensuremathTEMP{\mbox{\large$|$}}\:}

\newenvironment{block}{\begin{list}{\hbox{}}{\leftmargin 1em
    \itemindent -1em \topsep 0pt \itemsep 0pt \partopsep
    0pt}}{\end{list}}

\makeatletter \def\@listI{\leftmargin\leftmargini \parsep 4.5pt plus
1pt minus 1pt\topsep 6pt plus 2pt minus 2pt \itemsep 2pt plus 2pt
minus 1pt}

\let\@listi\@listI \@listi \makeatother

\makeatletter %
 \newcommand{\setoffdisplay}{\rule{5.9in}{1pt}}

\makeatother


\newcommand{\prob}{{\rm Prob}}

\newcommand{\numacc}{\ensuremathTEMP{{\rm \# acc}}}

\newcommand{\spp}{\ensuremathTEMP{{\rm SPP}}}

\newcommand{\rp}{{\rm RP}}

\newcommand{\nptm}{\rm NPTM}

\newcommand{\aor}{{\mathcal{A}}}

\newcommand{\oor}{{\mathcal{O}}}

\newcommand{\uap}{{\rm UAP}}

\newcommand{\auph}{{\rm AUPH}}
\newcommand{\ausigmap}[1]{\ensuremath{{{\rm AU}}\Sigma^{p}_{#1}}}

\newcommand{\aupip}[1]{\ensuremath{{{\rm AU}}\Pi^{p}_{#1}}}
\newcommand{\uph}{{\rm UPH}}
\newcommand{\usigmap}[1]{\ensuremath{{{\rm U}}\Sigma^{p}_{#1}}}

\newcommand{\upip}[1]{\ensuremath{{{\rm U}}\Pi^{p}_{#1}}}
\newcommand{\pruph}{{\rm \mathcal{U}\mathcal{P}\mathcal{H}}}
\newcommand{\prusigmap}[1]{\ensuremath{\mathcal{U}\Sigma^{p}_{#1}}}
\newcommand{\coprusigmap}[1]{\ensuremath{{{\rm co}}\mathcal{U}\Sigma^{p}_{#1}}}
\newcommand{\prupip}[1]{\ensuremath{\mathcal{U}\Pi^{p}_{#1}}}
\newcommand{\confl}{{\rm conflicting}}
\newcommand{\sens}{{\rm sensitive}}
\newcommand{\mlor}{{\mathcal{L}}}
\newcommand{\help}{{\rm help}}
\newcommand{\promiseup}{{\rm Promise\hbox{-}\up}}
\newcommand{\gi}{{\rm GI}}
\newcommand{\acc}{{\rm acc}}

\newcommand\INDXX{XX}

\newcommand\TABSET{999 \= \INDXX \= \INDXX \= \INDXX \= \INDXX \= \INDXX \= \INDXX \= \INDXX \= \INDXX \= \INDXX \= \INDXX \kill}

\begin{document}

\title{Hierarchical Unambiguity\footnote{A preliminary version of this paper
was presented at the MFCS~'06 conference.}}
\author{
\emph{Holger Spakowski}\footnote{Supported in part by the DFG under grants RO 1202/9-1 and RO 1202/9-3.}\\
    Institut f\"{u}r Informatik\\
  Heinrich-Heine-Universit\"{a}t D\"{u}sseldorf\\
  40225 D\"{u}sseldorf, Germany\\
  {\protect\tt{}spakowsk@}\protect\linebreak[0]\mbox{\protect\tt{}cs.uni-duesseldorf.de}
\and
\emph{Rahul Tripathi} \\
  Department of Computer Science and Engineering\\
  University of South Florida\\
  Tampa, FL 33620, USA\\
  {\protect\tt{}tripathi@}\protect\linebreak[0]\mbox{\protect\tt{}cse.usf.edu}\\
}

\date{}

\typeout{WARNING: BADNESS used to supress reporting.  Beware.}
\hbadness=3000%
\vbadness=10000 %

\bibliographystyle{alpha}
\pagestyle{empty}

\maketitle

\begin{abstract}
We develop techniques to investigate relativized hierarchical
unambiguous computation. We apply our techniques to generalize known
constructs involving relativized unambiguity based complexity
classes ($\up$ and $\promiseup$) to new constructs involving
arbitrary higher levels of the relativized unambiguous polynomial
hierarchy ($\uph$). Our techniques are developed on constraints
imposed by hierarchical arrangement of \emph{unambiguous}
nondeterministic polynomial-time Turing machines, and so they differ
substantially, in applicability and in nature, from standard methods
(such as the switching lemma~\cite{has:thesis:small-depth}), which
play roles in carrying out similar generalizations.

Aside from achieving these generalizations, we resolve a question
posed by Cai, Hemachandra, and
Vysko\v{c}~\cite{cai-hem-vys:b:promise} on an issue related to
nonadaptive Turing access to $\up$ and adaptive smart Turing access
to $\promiseup$.
\end{abstract}

\medskip




\normalspacing

\pagestyle{plain}

\sloppy

\section{Introduction}\label{sec:introduction}

\subsection{Background}
\label{subsec:bckgrd}

\noindent Baker, Gill, and Solovay in their seminal
paper~\cite{bak-gil-sol:j:rel} introduced the concept of
relativization in complexity theory, and showed that the bottom
levels of the polynomial hierarchy $\p$ and $\np$ separate in some
relativized world. Baker and Selman~\cite{bak-sel:j:step2} made
progress in extending this relativized separation to the next levels
of the polynomial hierarchy: They proved that there is a relativized
world where $\Sigma^{p}_{2} \neq \Pi^{p}_{2}$.
However, Baker and Selman~\cite{bak-sel:j:step2} noted that their
proof techniques do not apply %
at higher levels of the polynomial hierarchy because of certain
constraints in their counting argument.
Thus, it required the development of entirely different proof
techniques for separating all the levels of the relativized
polynomial hierarchy. The landmark paper by Furst, Saxe, and
Sipser~\cite{fur-sax-sip:j:parity} established the connection
between the relativization of the polynomial hierarchy and lower
bounds for small depth circuits computing certain functions.
Techniques for proving such lower bounds were developed in a series
of
papers~\cite{fur-sax-sip:j:parity,sip:c:Borel,yao:c:separating,has:thesis:small-depth},
which were motivated by questions about the relativized structure of
the polynomial hierarchy. Yao~\cite{yao:c:separating} finally
succeeded in separating the levels of the relativized polynomial
hierarchy by applying these new techniques.
H{\aa}stad~\cite{has:thesis:small-depth} gave the most refined
presentation of these techniques via the \emph{switching lemma}.
Even to date, H{\aa}stad's switching
lemma~\cite{has:thesis:small-depth} is used as an essential tool to
separate relativized hierarchies, composed of classes stacked one on
top of another. (See, for
instance,~\cite{has:thesis:small-depth,ko:j:exact,ber-ulf:j:ppph-heirarchy,spa-rah:j:unambig-alternating}
where the switching lemma is used as a strong tool for proving the
feasibility of oracle constructions.)

A major contribution of our paper lies in demonstrating that known
oracle constructions involving the initial levels of the unambiguous
polynomial hierarchy ($\uph$) and the promise unambiguous polynomial
hierarchy ($\pruph$), i.e.~$\up$ and $\p^{\promiseup}_{s}$,
respectively, can be extended to oracle constructions involving
arbitrary higher levels of $\uph$ by application only of pure
counting arguments. In fact, it seems implausible to achieve these
extensions by well-known techniques from circuit complexity (e.g.,
the switching lemma~\cite{has:thesis:small-depth} and the polynomial
method surveyed
in~\cite{beigel:c:polynomial-circuit,reg:b:polynomial-technique-survey}).
The class $\up$ is the unambiguous version of $\np$. $\up$ has
proved to be useful for instance in studying worst-case one-to-one
one-way
functions~\cite{ko:j:operators,gro-sel:j:complexity-measures},
obtaining potential counterexamples to the Berman-Hartmanis
isomorphism conjecture~\cite{jos-you:j:kcreative}, and in the study
of the complexity of closure properties of
$\sharpp$~\cite{hem-ogi:j:closure}. Lange and
Rossmanith~\cite{lan-ros:c:up-circuit-and-hierarchy} generalized the
notion of unambiguity to higher levels of the polynomial hierarchy.
They introduced the following unambiguity based hierarchies:
$\auph$, $\uph$, and $\pruph$.
It is known that $\auph$ $\subseteq$ $\uph$ $\subseteq$ $\pruph$
$\subseteq$
$\uap$~\cite{lan-ros:c:up-circuit-and-hierarchy,cra-gla-regan-samik:c:protocol},
where $\uap$ (unambiguous alternating polynomial-time) is the analog
of $\up$ for alternating polynomial-time Turing machines. These
hierarchies received renewed interests in some recent papers (see,
for
instance,~\cite{aida-cras-regan-wat:j:unique-games,cra-gla-regan-samik:c:protocol,spa-rah:j:unambig-alternating,gla-tra:t:machines}).
Spakowski and Tripathi~\cite{spa-rah:j:unambig-alternating},
developing on circuit complexity-theoretic proof techniques of Sheu
and Long~\cite{lon-she:j:up-low-high}, and of Ko~\cite{ko:j:exact},
obtained results on the relativized structure of these hierarchies.
Spakowski and Tripathi~\cite{spa-rah:j:unambig-alternating} proved
that there is a relativized world where these hierarchies are
infinite. They also proved that for each $k \geq 2$, there is a
relativized world where these hierarchies collapse so that they have
exactly $k$ distinct levels and their $k$'th levels collapse to
$\pspace$. The present paper supplements this investigation with a
focus on the structure of the unambiguous polynomial hierarchy.

\subsection{Results}

\noindent We prove a combinatorial lemma
(Lemma~\ref{lemma:weakness-uph-machines}) and demonstrate its
usefulness in generalizing known relativization results involving
classes such as $\up$ and $\promiseup$ to new relativization results
that involve arbitrary levels of the unambiguous polynomial
hierarchy ($\uph$).

In Subsection~\ref{subsec:sep-pruph-uph}, we use
Lemma~\ref{lemma:weakness-uph-machines} to construct relativized
worlds in which certain inclusion relationships between bounded
ambiguity classes ($\up_{O(1)}$ and $\fewp$) and the levels of the
unambiguous polynomial hierarchy ($\uph$) do not hold.
Theorem~\ref{thm:up-leqk-notin-usigmapk} of this subsection subsumes
an oracle result of Beigel~\cite{bei:c:up1} for any constant $k \geq
1$ and Corollary~\ref{cor:k-turing-promiseup} generalizes a result
of Cai, Hemachandra, and Vysko\v{c}~\cite{cai-hem-vys:b:promise}
from the case of $k=2$ to the case of any arbitrary $k \geq 2$.

Subsection~\ref{subsec:nonadap-adap-nonpromise} studies the issue of
simulating nonadaptive access to $\usigmap{h}$, the $h$'th level of
the unambiguous polynomial hierarchy, by adaptive access to
$\usigmap{h}$. Theorem~\ref{thm:truth-table-versus-Turing} of this
subsection generalizes a result of Cai, Hemachandra, and
Vysko\v{c}~\cite{cai-hem-vys:c-OUT-BY-BOOK:promise} from the case of
$h=1$ to the case of any arbitrary $h \geq 1$.
Lemma~\ref{lemma:weakness-uph-machines} is used as a key tool for
proving Theorem~\ref{thm:truth-table-versus-Turing}.

We improve upon Theorem~\ref{thm:truth-table-versus-Turing} of
Subsection~\ref{subsec:nonadap-adap-nonpromise} in
Subsection~\ref{subsec:nonadap-adap-promise}. There are compelling
reasons for the transition from
Subsection~\ref{subsec:nonadap-adap-nonpromise} to
Subsection~\ref{subsec:nonadap-adap-promise}, which we discuss in
Subsection~\ref{subsec:nonadap-adap-promise}.
Theorem~\ref{thm:promis-truth-table-versus-Turing} in that
subsection not only resolves a question posed by Cai, Hemachandra,
and Vysko\v{c}~\cite{cai-hem-vys:b:promise}, but also generalizes
one of their results. In particular,
Theorem~\ref{thm:promis-truth-table-versus-Turing} holds for any
total, polynomial-time computable and polynomially bounded function
$k(\cdot)$ and arbitrary $h \geq 1$, while a similar result of Cai,
Hemachandra, and Vysko\v{c}~\cite{cai-hem-vys:b:promise} holds only
for any arbitrary \emph{constant} $k$ and $h=1$.
Lemma~\ref{lemma:weakness-uph-machines} is one of the ingredients in
the proof of this theorem.

Subsection~\ref{subsec:adap-nonadap} investigates the complimentary
issue of simulating adaptive access to $\usigmap{h}$ by nonadaptive
access to $\usigmap{h}$. Theorem~\ref{thm:const-kh-Ttt} of this
subsection generalizes a result of Cai, Hemachandra, and
Vysko\v{c}~\cite{cai-hem-vys:b:promise} from the case of $h=1$ to
the case of any arbitrary constant $h \geq 1$. Again,
Lemma~\ref{lemma:weakness-uph-machines} is useful in making this
generalization possible.

In Subsection~\ref{subsec:fault-tolerance}, we study the notion of
one-sided helping introduced by Ko~\cite{ko:j:helping}.
Corollary~\ref{cor:helping-up-k} of this subsection generalizes and
improves one of the results of Cai, Hemachandra, and
Vysko\v{c}~\cite{cai-hem-vys:b:promise}.

Finally, in Section~\ref{sec:robust-unambig-sigmak} we consider the
possibility of imposing a more stringent restriction in the
statement of Lemma~\ref{lemma:weakness-uph-machines}. The
investigation in this subsection leads to a generic oracle collapse
of $\uph$ to $\p$
under the assumption $\p = \np$. This extends a result of Blum and
Impagliazzo~\cite{blu-imp:c:generic}, which showed a generic oracle
collapse of $\up$ (the first level of $\uph$) to $\p$ assuming $\p =
\np$.

\section{Preliminaries}
\label{sec:prelim}

\subsection{Notations}
\label{subsec:notations}

\noindent Let $\naturalnumber^+$ denote the set of positive
integers. $\Sigma$ denotes the alphabet $\{0,1\}$. Let $[n] =_{df} \{1, 2, \ldots, n\}$
for every $n \in \naturalnumber^+$. $\nptm$ stands
for ``nondeterministic polynomial-time Turing machine." For every
oracle $\nptm$ $N$, oracle $A$, and string $x \in \sigmastar$, we
use the shorthand $N^{A}(x)$ for ``the computation tree
of $N$ with oracle $A$ on input $x$." We fix a standard,
polynomial-time computable and invertible, one-to-one, multiarity
pairing function $\langle ., \ldots, .\rangle$ throughout the paper.
Let $\circ$ denote the composition operator on functions. For any
polynomial $p(.)$ and integer $i \geq 1$, let $(p \circ)^{i}(\cdot)$
denote $\underbrace{p \circ p \circ \cdots \circ p}_{i}(\cdot)$,
i.e.,~the polynomial obtained by $i$ compositions of $p$. All
polynomials $p(\cdot)$ appearing in this paper are without loss of
generality nondecreasing and satisfy $p(n) \geq n$ for every $n \in
\naturalnumber^{+}$. Let $\sigma$ be an equivalence relation on a
set $S$. For each $x \in S$, the \emph{equivalence class} [x] of $x$
determined by $\sigma$ is $\{y \in S~|~x \sigma y\}$. The set
$S/\sigma$ of all equivalence classes determined by $\sigma$ is
called the {\em quotient set} determined by $\sigma$. For any set
$S$, we use $\powerset(S)$ to denote the \emph{power set} of $S$,
i.e., the set of all subsets of $S$. The \emph{join} of two sets $A$
and $B$ over $\Sigma$ is defined as $A \oplus B = \{ 0x \condition
x\in A\} \cup \{ 1x \condition x\in B\}$.

We define the notion of computation path of oracle machines
independent of any concrete oracle. A {\em computation path} of an
oracle $\nptm$ $N$ encodes a complete valid computation that $N$ can
have relative to some/any oracle, i.e., it contains the sequence of
configurations including the query strings and the answers from the
oracle. Hence two computation paths $\rho_1$ and $\rho_2$ of an
oracle NPTM are equal if and only if the configuration sequences,
oracles queries, and oracles answers are the same for the
computation paths. For any computation path $\rho$, let $Q^+(\rho)$
denote the set of strings that are queried along $\rho$ and answered
positively, and let $Q^-(\rho)$ denote the set of strings that are
queried along $\rho$ and answered negatively. Let $Q(\rho) =
Q^+(\rho) \cup Q^-(\rho)$. For any concrete oracle $A$ and input
$x$, a given path $\rho$ may or may not appear
in $N^A(x)$. For instance, if $\alpha \in Q^+(\rho)$ then $\rho$
does not appear in $N^A(x)$ for any $A$ with $\alpha \notin A$. In
this case we also say ``$N^A(x)$ does not have path $\rho$.''

For any complexity class $\mathcal{C}$ and for any
natural notion of polynomial-time reducibility $r$ (e.g., $r \in
\{m, dtt, tt, k\hbox{-}tt, T, k\hbox{-}T, b\}$), let
$R^{p}_{r}(\mathcal{C})$ denote the closure of $\mathcal{C}$ under
$r$. That is, $R^{p}_{r}(\mathcal{C}) =_{df} \{L~|~(\exists L' \in
\mathcal{C})[L \leq^{p}_{r} L']\}$. We refer the reader to any
standard textbook in complexity theory
(e.g.~\cite{bov-cre:b:complexity,hem-ogi:b:companion,pap:b:complexity}) for complexity classes and reductions not defined
in this paper.

Given a complexity class $\mathcal{C}$, the unique existential
$(\exists!)$ and the unique universal $(\forall!)$ operators on
$\mathcal{C}$ yield complexity classes. Formally:

\begin{definition}
\label{def:unique-quantifiers} For any arbitrary complexity class
$\mathcal{C}$,
\begin{enumerate}
\item $\exists ! \cdot \mathcal{C}$ is defined to be the class of all sets $L$
for which there exists a polynomial $p(\cdot)$ and a set $L' \in
\mathcal{C}$ such that for all $x \in \sigmastar$,
\begin{eqnarray*}
x \in L &\Longrightarrow& (\mbox{there exists a unique $y \in
\Sigma^{p(|x|)}$})[\langle x, y\rangle \in L'], \mbox{and} \\
x \not \in L &\Longrightarrow& (\mbox{for all $y \in
\Sigma^{p(|x|)}$})[\langle x, y\rangle \not \in L'].
\end{eqnarray*}

\item $\forall !\cdot \mathcal{C}$ is defined to be the class of all sets $L$
for which there exists a polynomial $p(\cdot)$ and a set $L' \in
\mathcal{C}$ such that for all $x \in \sigmastar$,
\begin{eqnarray*}
x \in L &\Longrightarrow& (\mbox{for all $y \in \Sigma^{p(|x|)}$})[\langle x, y\rangle \in L'], \mbox{and} \\
x \not \in L &\Longrightarrow& (\mbox{there exists a unique $y \in
\Sigma^{p(|x|)}$})[\langle x, y\rangle \not \in L'].
\end{eqnarray*}

\end{enumerate}
\end{definition}

\noindent We introduce the notion of a $\Sigma_{k}(A)$-system. This
notion is useful for concisely representing the computation of a
stack of oracle $\nptm$s.

\begin{definition}
\label{def:sigmakA-system}
\begin{enumerate}
\item For any $k \in \naturalnumber^{+}$ and $A \subseteq \sigmastar$,
we call a tuple $[A; N_1, N_2, \ldots, N_k]$, where $A$ is an oracle
and $N_1, N_2, \ldots, N_k$ are nondeterministic oracle Turing
machines, a $\Sigma_{k}(A)$-system. The {\em computation of a
$\Sigma_{k}(A)$-system} $[A; N_1, N_2, \ldots, N_k]$ on input $x$,
denoted by $[A; N_1, N_2, \ldots, N_k](x)$, is defined as follows:
\begin{itemize}
\item For $k=1$, $[A;N_1](x) =_{df} N_{1}^{A}(x)$, and
\item for $k>1$, $[A; N_{1}, N_{2}, \ldots, N_{k}](x) =_{df}
N_{1}^{L(N_{2}^{\cdot^{\cdot^{\cdot^{L(N_{k}^{A})}}}})}(x)$.
\end{itemize}
\item
{\em The language accepted by a $\Sigma_{k}(A)$-system}, denoted by $L[A;
N_1, N_2, \ldots, N_k]$, is defined inductively as follows:
\[
L[A; N_1, N_2, \ldots, N_k] =_{df} \left\{
\begin{array}{ll}
L(N_{1}^{A}) & \mbox{if $k=1$, and} \\
L(N_{1}^{L[A; N_2, N_3, \ldots, N_{k}]}) & \mbox{if $k > 1$}.
\end{array}
\right.
\]
\end{enumerate}
\end{definition}

\noindent We define the notion of unambiguity in
$\Sigma_{k}(A)$-systems as follows:

\begin{definition}
\label{def:unambig-sigmakA}
\begin{enumerate}
\item
We say that a $\Sigma_{k}(A)$-system $[A; N_1, N_2, \ldots,
N_k]$ is {\em unambiguous} if for every $1 \leq i \leq k$ and for every $x
\in \sigmastar$, $[A; N_i, N_{i+1}, \ldots, N_k](x)$ has at most one
accepting path.
\item
For any $\Sigma_{k}(A)$-system $[A; N_1, N_2, \ldots, N_k]$, we
define
\[
L_{\textnormal{unambiguous}}[A; N_1, N_2, \ldots, N_k] = \left\{
\begin{array}{ll}
L[A; N_1, N_2, \ldots, N_k] & \mbox{if } [A; N_1, N_2, \ldots, N_k] \mbox{ is } \\
&                             \mbox{unambiguous}, \\
\mbox{undefined} & \mbox{ otherwise}.
\end{array}
\right.
\]
\end{enumerate}
\end{definition}

\noindent Roughly speaking, a property of an oracle machine is
called \emph{robust} if the machine retains that property with
respect to every oracle. Below we define the property of
\emph{robust unambiguity} for a $\Sigma_k(A)$-system.

\begin{definition}
\label{def:robust-unambig-sigmakA}
We say that a $\Sigma_k(A)$-system $[A; N_1, N_2, \ldots, N_k]$ is
\emph{robustly unambiguous} if for every set $B$, the
$\Sigma_{k}(A\oplus B)$-system $[A \oplus B; N_1, N_2, \ldots, N_k]$
is unambiguous.
\end{definition}

\subsection{Promise Problems and Smart Reductions}
\label{subsec:promise-problem}

\noindent Even, Selman, and
Yacobi~\cite{eve-sel-yac:j:promise-problems} introduced and studied
the notion of promise problems. Promise problems are generalizations
of decision problems in that the set of Yes-instances and the set of
No-instances must partition the set of all instances in a decision
problem, whereas this is not necessarily the case with promise
problems. Thus, for a promise problem a set of disallowed strings
may be defined, which represent neither Yes-instances nor
No-instances. Over the years, the notion of promise problems has
proved to be useful at several places in computational complexity
theory. (See~\cite{gol:t:promise-problem} for a nice survey on some
applications of promise problems in computational complexity
theory.)

\begin{definition}[Based on~\cite{gol:t:promise-problem};
cf.~\cite{eve-sel-yac:j:promise-problems}] A promise problem $\Pi =
(\Pi_{\textnormal{yes}}, \Pi_{\textnormal{no}})$ is defined in terms
of disjoint sets $\Pi_{\textnormal{yes}}$, $\Pi_{\textnormal{no}}
\subseteq \sigmastar$. The set $\Pi_{\textnormal{yes}}$ is called
the set of Yes-instances, the set $\Pi_{\textnormal{no}}$ is called
the set of No-instances, and the set $\Pi_{\textnormal{yes}} \cup
\Pi_{\textnormal{no}}$ is called the promise set.
\end{definition}

\noindent Some technicalities are involved when oracle access to a
promise problem is defined. If a query to a promise problem falls
inside the promise set, then the answer to the query is well-defined
(i.e., the answer is $1$ if $q$ is a Yes-instance and is $0$ if $q$
is a No-instance). However, if a query falls outside the promise
set, then it is not immediately clear how that query should be
handled by the promise problem, i.e., the oracle. Several natural
models of oracle access to a promise problem are definable.
(See~\cite{gro-sel:j:complexity-measures,cai-hem-vys:b:promise} for
a few possible approaches to defining oracle accesses to promise
problems.)

Grollmann and Selman~\cite{gro-sel:j:complexity-measures} proposed a
model of oracle access to a promise problem that {\em prohibits}
queries that fall outside the promise set. In this model, a querying
machine always asks queries from the promise set, i.e., the queries
asked by the querying machine always obey the underlying promise of
the promise problem. For instance, let us define a promise problem
$\Pi_{\textnormal{unique}} = (\Pi_{\textnormal{yes}},
\Pi_{\textnormal{no}})$ in terms of acceptance mechanism of a
$\nptm$ $N$ as follows: $\Pi_{\textnormal{yes}} = \{x\in
\sigmastar~|~\numacc_{N}(x) = 1\}$ and $\Pi_{\textnormal{no}} =
\{x\in \sigmastar~|~\numacc_{N}(x) = 0\}$. Then a Turing access to
$\Pi_{\textnormal{unique}}$ in the model proposed by Grollmann and
Selman~\cite{gro-sel:j:complexity-measures} requires that for any
query $y$ asked by the querying machine on some input, the
computation of $N$ on $y$ must be unambiguous, i.e.,
$\numacc_{N}(y)$ must be either $0$ or $1$. A Turing reduction that
obeys the constraints of this model (i.e., any query ever asked
belongs to the promise set) is called a \emph{smart} Turing
reduction~\cite{gro-sel:j:complexity-measures}. The definition given
below formally captures the notion of smart Turing reduction from a
decision problem to a promise problem.\footnote{Cai, Hemachandra,
and Vysko\v{c}~\cite{cai-hem-vys:b:promise} referred to Grollmann
and Selman's \emph{smart} oracle access by the term \emph{guarded}
access.}

\begin{definition}
A set $L$ polynomial-time \emph{smart} Turing reduces to a promise
problem $\Pi = (\Pi_{\textnormal{yes}}, \Pi_{\textnormal{no}})$,
denoted by $L \leq^{p}_{s, T} \Pi$ or $L \in \p_{s}^{\Pi}$, if there
is a deterministic polynomial-time oracle Turing machine $M$ such
that for all $x \in \sigmastar$,
\begin{enumerate}
\item $x \in L \Longleftrightarrow M^{\Pi}(x)$ accepts, and
\item if $M^{\Pi}(x)$ asks a query $y$ to $\Pi$, then $y \in \Pi_{\textnormal{yes}} \cup
\Pi_{\textnormal{no}}$.
\end{enumerate}
If on all inputs $x \in \sigmastar$, the querying machine $M$ asks
at most $k$ queries, for some integer constant $k \geq 1$, then we
say that $L$ polynomial-time \emph{smart} $k$-Turing reduces to
$\Pi$ and write $L \leq^{p}_{s, k\hbox{-}T} \Pi$ or $L \in
\p_{s}^{\Pi[k]}$.
\end{definition}

\noindent In the above definition, we followed Grollmann and
Selman's notion of smart Turing reductions from \emph{decision}
problems to \emph{promise} problems. We may extend this notion to
define reductions that reduce \emph{promise} problems to
\emph{promise} problems. (See, for
instance,~\cite{gol:t:promise-problem} for a generalization of smart
Turing reductions to reductions among promise problems.) In this
paper, we will only consider smart Turing reductions (i.e.,
reductions from decision problems to promise problems) as given by
Grollmann and Selman.

The following two definitions are standard.

\begin{definition}
Let $\Pi$ be any promise problem. $R^{p}_{s,T}(\Pi)$ is the
class of all sets $L$ such that $L \leq^{p}_{s, T} \Pi$;\, for all
$k \in \naturalnumber^+$, $R^{p}_{s, k\hbox{-}T}(\Pi)$ is
the class of all sets $L$ such that $L \leq^{p}_{s, k\hbox{-}T}
\Pi$;\, $R^{p}_{s, b}(\Pi)$ is the class of all sets $L$
for which there exists some $k \in \naturalnumber^{+}$ such that $L
\leq^{p}_{s, k\hbox{-}T} \Pi$.
\end{definition}

\begin{definition}
For any class of promise problems $\mathcal{C}$ and any reduction
$r\in\{ T, k\hbox{-}T, b \}$, we define
$R^{p}_{s,r}(\mathcal{C}) =_{df} \bigcup_{\Pi \in \mathcal{C}}
R^{p}_{s,r}(\Pi)$.
\end{definition}

\noindent We will study the computational power of smart Turing
reductions to a particular class of promise problems, namely
the class $\promiseup$, which is defined as follows.
\begin{definition}
$\promiseup$ is the class of all promise problems $\Pi =
(\Pi_{\textnormal{yes}}, \Pi_{\textnormal{no}})$ for which there
exists a nondeterministic polynomial-time Turing machine $N$ such
that for all $x \in \sigmastar$,
\begin{eqnarray*}
x \in \Pi_{\textnormal{yes}} &\Longrightarrow& \numacc_{N}(x) = 1,
\mbox{ and } \\
x \in \Pi_{\textnormal{no}} &\Longrightarrow& \numacc_{N}(x) = 0.
\end{eqnarray*}
\end{definition}

\noindent
The class $\p_{s}^{\promiseup}$ of sets that
polynomial-time smart Turing reduce to $\promiseup$ is a prominent class
that behaves remarkably differently than the related class $\p^{\up}$. While $\p_{s}^{\promiseup}$
is known to contain the class $\fewp$ %
and the graph isomorphism problem~\cite{arvind-kurur:c:gi}, %
similar results for the case of $\p^{\up}$ are
unknown.\footnote{Arvind and Kurur~\cite{arvind-kurur:c:gi} showed
that the graph isomorphism problem $(\gi)$ belongs to $\spp$, a
class introduced
in~\cite{gup:j:closure-division,hem-ogi:j:closure,fen-for-kur:j:gap}.
Subsequently, Crasmaru et al.~\cite{cra-gla-regan-samik:c:protocol}
observed that the proof of classifying $\gi$ into $\spp$, as given
by Arvind and Kurur~\cite{arvind-kurur:c:gi}, actually yields a
somewhat improved classification for $\gi$: $\gi$ belongs to
$R^{p}_{s,T}(\promiseup)$, a subclass of
$\spp$~\cite{cra-gla-regan-samik:c:protocol}.}

\subsection{Unambiguity Based Hierarchies}
\label{subsec:unambig-hierarchy}

\noindent Niedermeier and Rossmanith~\cite{nie-ros:j:up-hierarchy}
observed that the notion of unambiguity in $\nptm$s can be
generalized in three ways, each of which define an unambiguity based
hierarchy.

\begin{definition}[Unambiguity Based Hierarchies~\cite{lan-ros:c:up-circuit-and-hierarchy,nie-ros:j:up-hierarchy}]
\label{def:up-hierarchy}
\begin{enumerate}
\item The \emph{alternating unambiguous polynomial hierarchy} is
defined as:
\[
\auph =_{df} \bigcup_{k \geq 0} \ausigmap{k} = \bigcup_{k \geq 0}
\aupip{k},
\]
where
\[
\ausigmap{k} = \left\{
\begin{array}{ll}
\p & \mbox{if $k=0$,} \\
\exists! \cdot \aupip{k-1} & \mbox{if $k \geq 1$,}
\end{array}
\right. \mbox{\hspace*{1 cm} and \hspace*{1 cm}} \aupip{k} = \left\{
\begin{array}{ll}
\p & \mbox{if $k=0$,} \\
\forall! \cdot \ausigmap{k-1} & \mbox{if $k \geq 1$.}
\end{array}
\right.
\]
\item The \emph{unambiguous polynomial hierarchy} is defined as:
\[
\uph =_{df} \bigcup_{k \geq 0} \usigmap{k} = \bigcup_{k \geq 0}
\upip{k},
\]
where
\[
\usigmap{k} = \left\{
\begin{array}{ll}
\p & \mbox{if $k=0$,} \\
\up^{\usigmap{k-1}} & \mbox{if $k \geq 1$,}
\end{array}
\right. \mbox{\hspace*{1 cm} and \hspace*{1 cm}} \upip{k} = \left\{
\begin{array}{ll}
\p & \mbox{if $k=0$,} \\
\coup^{\usigmap{k-1}} & \mbox{if $k \geq 1$.}
\end{array}
\right.
\]
\item The \emph{promise unambiguous polynomial hierarchy} is defined as:
\[
\pruph =_{df} \bigcup_{k \geq 0} \prusigmap{k} = \bigcup_{k \geq 0}
\prupip{k},
\]
where $\prusigmap{0} =_{df} \p$, $\prusigmap{1} =_{df} \up$, and for
every $k \geq 2$, $\prusigmap{k}$ is the class of all sets $L \in
\Sigma^{p}_{k}$ such that for some oracle $\nptm$s $N_1$, $N_2$,
$\ldots$, $N_k$, $L =
L(N_{1}^{L(N_{2}^{\cdot^{\cdot^{\cdot^{L(N_{k})}}}})})$, and for
every $x \in \sigmastar$ and for every $1 \leq i \leq k-1$,
$N_{1}^{L(N_{2}^{\cdot^{\cdot^{\cdot^{L(N_{k})}}}})}(x)$ has at most
one accepting path and if $N_{i}$ asks a query $w$ to its oracle
$L(N_{i+1}^{\cdot^{\cdot^{\cdot^{L(N_k)}}}})$ during the computation
of $N_{1}^{\cdot^{\cdot^{\cdot^{L(N_k)}}}}(x)$, then
$N_{i+1}^{\cdot^{\cdot^{\cdot^{L(N_k)}}}}(w)$ has at most one
accepting path. For each $k \geq 0$, the class $\prupip{k}$ is
defined to be: $\prupip{k} =_{df} \coprusigmap{k}$.
\end{enumerate}
\end{definition}

\begin{figure}[t]
\begin{center}
\includegraphics[scale=0.6]{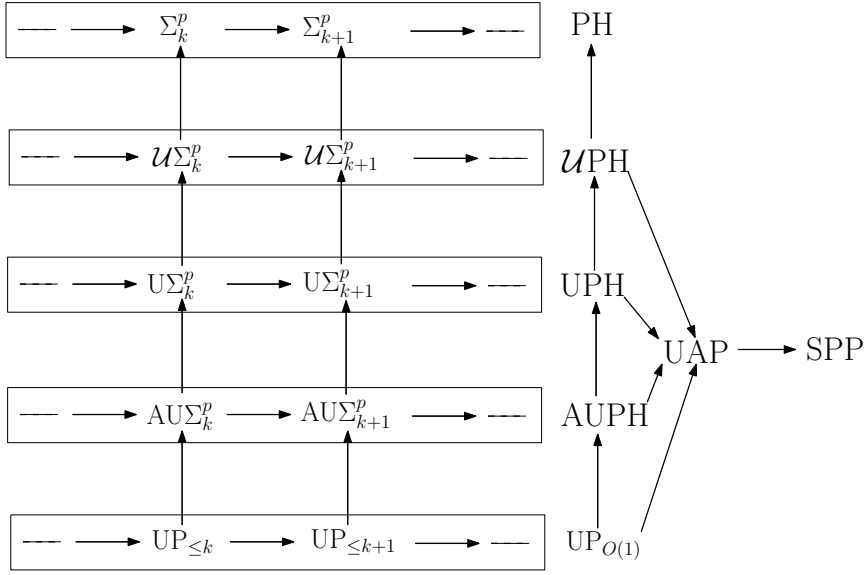}

\end{center}
\caption{Known inclusion structure of unambiguity based classes and
other central classes. The arrows point from subclasses to
superclasses.} \label{fig:complexity-relation}
\end{figure}

\noindent The following inclusion relationships between unambiguity
based classes and other central classes are known (see also
Figure~\ref{fig:complexity-relation}).

\begin{theorem}
\label{thm:summary-unamb}
\begin{enumerate}
\item For all $k \geq 0$, $\ausigmap{k} \subseteq \usigmap{k} \subseteq \prusigmap{k} \subseteq \Sigma^{p}_{k}$~\textnormal{\cite{lan-ros:c:up-circuit-and-hierarchy}}.
\item For all $k \geq 1$, $\up_{\leq k} \subseteq \ausigmap{k} \subseteq \usigmap{k} \subseteq \prusigmap{k} \subseteq \uap \subseteq \spp$~\textnormal{(\cite{lan-ros:c:up-circuit-and-hierarchy}
+~\cite{nie-ros:j:up-hierarchy}
+~\cite{cra-gla-regan-samik:c:protocol})}.
\end{enumerate}
\end{theorem}

\noindent Despite the attention these hierarchies deserve, much less
is known about the structure of these hierarchies since Lange and
Rossmanith~\cite{lan-ros:c:up-circuit-and-hierarchy} first posed
questions---such as, whether these hierarchies intertwine, or
whether some unambiguity based hierarchy is contained in a fixed
level of some other hierarchy, or whether some/all of these
hierarchies collapse to a fixed level---on their structure. On the
positive side, there had been some advances in understanding the
structure of these hierarchies. Hemaspaandra and
Rothe~\cite{hem-rot:j:boolean} related the structure of these
hierarchies to the existence of sparse Turing complete sets for
$\up$. The structure of these hierarchies received renewed interests
in some recent works
(see~\cite{aida-cras-regan-wat:j:unique-games,cra-gla-regan-samik:c:protocol,spa-rah:j:unambig-alternating,gla-tra:t:machines}).
In particular, Spakowski and
Tripathi~\cite{spa-rah:j:unambig-alternating} investigated the
relativized structure of these hierarchies. They proved that the
unambiguity based hierarchies $\auph$, $\uph$, and $\pruph$ are
infinite in some relativized world. They also proved a contrasting
result on their relativized structure: For each $k \geq 2$, there is
a relativized world where these hierarchies collapse so that they
have exactly $k$ distinct levels and their $k$'th levels coincide
with $\pspace$.

\section{Proof Technique}

\subsection{Main Lemma}

\noindent Our main lemma is Lemma~\ref{lemma:weakness-uph-machines},
which we will use throughout this paper for generalizing known
oracle constructions involving unambiguity based classes such as
$\up$ and $\promiseup$ to new oracle constructions involving
arbitrary levels of the $\uph$. Roughly,
Lemma~\ref{lemma:weakness-uph-machines} states computational
limitations of a $\Sigma_{k}(\oor)$-system, for any arbitrary $k
\geq 1$, under certain weak conditions.

\begin{lemma}
\label{lemma:weakness-uph-machines}
Fix a $\Sigma_{k} (\oor)$-system $[\oor; N_1, N_2, \ldots, N_k]$,
a string $x \in \sigmastar$, and a set $U \subseteq \sigmastar$ such that
$\oor \cap U = \emptyset$. %
Let $r(.)$ be a polynomial that bounds the running time of each of
the machines $N_1, N_2, \ldots, N_k$. Then the following holds:
\begin{enumerate}

\item Suppose $[\oor; N_1, N_2, \ldots, N_k](x)$ accepts and
for every $A \subseteq U$ with $||A|| \leq k$, $[\oor \cup A; N_1,
N_2, \ldots, N_k]$ is unambiguous. Let
\[
C = \{\alpha \in U~|~[\oor \cup \{\alpha\}; N_1, N_2, \ldots,
N_k](x) \mbox{ rejects}\}.
\]
Then $||C|| \leq 5^{k} \cdot \prod_{i=1}^{k}(r \circ)^{i}(|x|)$.

\item Suppose $[\oor; N_1, N_2, \ldots, N_k](x)$ rejects and
for every $A \subseteq U$ with $||A|| \leq k+1$, $[\oor \cup A; N_1,
N_2, \ldots, N_k]$ is unambiguous. Let
\[
C = \{\alpha \in U~|~[\oor \cup \{\alpha\}; N_1, N_2, \ldots,
N_k](x) \mbox{ accepts}\}.
\]
Then $||C|| \leq 5^{k} \cdot \prod_{i=1}^{k}(r \circ)^{i}(|x|)$.

\end{enumerate}
\end{lemma}

\sproof We prove $(1)$ and $(2)$ by induction on $k$. For the base
case $k=1$ of $(1)$, we have $[\oor; N_1](x)$ accepts. Also since
$[\oor; N_1]$ is unambiguous by the assumption made in (1),
there is a unique accepting path in $[\oor; N_1](x)$. Let $C'$ be
the set of all queries $w \in U$ along this unique accepting path.
Then clearly $||C|| \leq ||C'|| \leq r(|x|)$. Thus $(1)$ holds for
the base case.

For the base case $k=1$ of $(2)$, we have $[\oor; N_{1}](x)$
rejects. Suppose that $||C|| > 5 \cdot r(|x|)$. Note that for every
$\alpha \in C$, $[\oor \cup \{\alpha\}; N_1](x)$ accepts.
Thus, for every $\alpha \in C$, let $\lambda(\alpha)$ be an
accepting path in $[\oor \cup \{\alpha\}; N_1](x)$. It is easy to
show that $\lambda(\alpha_1) \not= \lambda(\alpha_2)$ for any
distinct $\alpha_1, \alpha_2 \in C$. To see this, let $\rho
=\lambda(\alpha)$ for some $\alpha\in C$. Then, by the definition of
$\lambda(\alpha)$ and the assumption that $[\oor; N_1](x)$ rejects,
we notice that path $\rho$ appears in $N_1^{\oor \cup \{ \alpha
\}}(x)$ but does not appear in $N_1^{\oor}(x)$. From this, it
follows that $\alpha$ must be answered positively along $\rho$,
i.e., $\alpha \in Q^+(\rho)$, since otherwise $\rho$ would also
appear in $N_1^{\oor }(x)$. Therefore, for any oracle $\mathcal{B}$
with $\alpha \notin\mathcal{B}$, $\rho$ cannot appear in
$N_1^{\mathcal{B}}(x)$. In particular, $\rho$ cannot appear in
$N_1^{ \oor \cup \{ \alpha' \} }(x)$ for any $\alpha'$ with $\alpha'
\not= \alpha$. Hence, we get that $\lambda(\alpha_1)
\not=\lambda(\alpha_2)$ for any distinct $\alpha_1, \alpha_2 \in C$.

We define for any $\alpha \in C$,
\begin{eqnarray*}
\confl(\alpha) = \{\beta \in C~|~\mbox{$\lambda(\alpha)$ is not an
accepting path in $[\oor \cup \{\alpha, \beta\}; N_1]$}\}.
\end{eqnarray*}

\noindent Since $N_{1}(x)$ with any oracle asks at most $r(|x|)$
queries, there can be at most $r(|x|)$ strings $\beta \in C$ that
can cause $\lambda(\alpha)$ not to appear in $[\oor \cup \{\alpha,
\beta\}; N_1]$. In other words, for any $\alpha \in C$, it holds
that $||\confl(\alpha)|| \leq r(|x|)$. Thus, it follows by an easy
counting argument and the assumption that $||C|| > 5 \cdot r(|x|)$
that there exist distinct $\alpha_1, \alpha_2 \in C$ such that
$\alpha_1 \not \in \confl(\alpha_2)$ and $\alpha_2 \not \in
\confl(\alpha_1)$. We had already shown that $\lambda(\alpha_1)$ and
$\lambda(\alpha_2)$ are distinct paths for distinct $\alpha_1,
\alpha_2 \in C$. Therefore, $[\oor \cup \{\alpha_1, \alpha_2\};
N_1](x)$ has two distinct accepting paths, namely
$\lambda(\alpha_1)$ and $\lambda(\alpha_2)$. This contradicts our
assumption that $[\oor \cup A; N_1]$ is unambiguous for every $A
\subseteq U$ with $||A|| \leq 2$. Thus also $(2)$ holds for the base case.\\

\noindent \textbf{Induction Hypothesis:} Assume that $(1)$ and $(2)$
in the statement of Lemma~\ref{lemma:weakness-uph-machines} hold
for some $k \geq 1$. \\

\noindent \textbf{Inductive Step:} Let $[\oor; N_1, N_2, \ldots,
N_{k+1}]$ be a $\Sigma_{k+1}(\oor)$-system and let $r(.)$ be a
polynomial that bounds the running time of each of $N_1, N_2,
\ldots, N_k, N_{k+1}$.

We first prove $(1)$. Suppose $[\oor; N_1, N_2, \ldots, N_{k+1}](x)$
accepts and $[\oor \cup A; N_1, N_2, \ldots, N_{k+1}]$ is
unambiguous for every $A \subseteq U$ with $||A|| \leq k+1$. Let
$\lambda$ denote the unique accepting path of $[\oor; N_1, N_2,
\ldots, N_{k+1}](x)$. For every query $w$ along $\lambda$, the
$\Sigma_{k}(\oor)$-system $[\oor; N_2, N_3, \ldots, N_{k+1}]$
computes $[\oor; N_2, N_3, \ldots, N_{k+1}](w)$. By
Definition~\ref{def:unambig-sigmakA}(1), for every $A \subseteq U$
with $||A|| \leq k+1$, the $\Sigma_{k}(A)$-system $[\oor \cup A;
N_2, N_3, \ldots, N_{k+1}]$ is unambiguous. Thus, it follows by the
induction hypothesis that for every query $w$ along $\lambda$ and
for all but at most $5^{k} \cdot \prod_{i=1}^{k}(r \circ)^{i}(|w|)
\leq 5^{k} \cdot \prod_{i=2}^{k+1}(r \circ)^{i}(|x|)$ strings
$\alpha \in U$, adding $\alpha \in U$ to $\oor$ does not change the
decision (i.e., acceptance or rejection) of $[\oor; N_2, N_3,
\ldots, N_{k+1}](w)$. Since the number of such queries $w$ along
$\lambda$ is at most $r(|x|)$, it follows that $\lambda$ is an
accepting path in $[\oor \cup \{\alpha\}; N_1, N_2, \ldots,
N_{k+1}](x)$ for all but at most $r(|x|) \cdot 5^{k} \cdot
\prod_{i=2}^{k+1}(r \circ)^{i}(|x|) < 5^{k+1} \cdot
\prod_{i=1}^{k+1}(r \circ)^{i}(|x|)$ strings $\alpha \in U$. This
proves $(1)$, the first part of the inductive step.

We now prove $(2)$. Suppose that $||C|| > 5^{k+1} \cdot \prod_{i=1}^{k+1}(r \circ)^{i}(|x|)$.
For any $\alpha \in C$, let $\lambda(\alpha)$ denote the unique
accepting path in $[\oor \cup \{\alpha\}; N_1, N_2, \ldots,
N_{k+1}](x)$. We define an equivalence relation $\sigma$ on $C$ as
follows: For every $\alpha_1, \alpha_2 \in C$,
\[
\alpha_1 \sigma \alpha_2 \Longleftrightarrow \lambda(\alpha_1) =
\lambda(\alpha_2).
\]
The following cases are exhaustive.

\medskip

\noindent {\boldmath \bf Case 1: There is an equivalence class of
$\sigma$ of size $> 5^{k} \cdot \prod_{i=1}^{k+1}(r
\circ)^{i}(|x|)$.} Let $[\alpha]$ be such an equivalence class. Then
we have the following situation: The accepting path
$\lambda(\alpha)$ does not appear in $[\oor; N_1, N_2, \ldots,
N_{k+1}](x)$, but for every $\beta\in [\alpha]$, path
$\lambda(\alpha)$ appears in $[\oor\cup \{ \beta \} ; N_1, N_2,
\ldots, N_{k+1}](x)$. Hence for every query $w\in\sigmastar$ along
$\lambda(\alpha)$, it holds that for every $\beta, \beta' \in
[\alpha]$, $[\oor\cup \{ \beta \} ; N_2, N_3, \ldots, N_{k+1}](w)$
and $[\oor\cup \{ \beta' \} ; N_2, N_3, \ldots, N_{k+1}](w)$ have
the same acceptance behavior, i.e, $[\oor\cup \{ \beta \} ; N_2,
N_3, \ldots, N_{k+1}](w)$ accepts if and only if $[\oor\cup \{
\beta' \} ; N_2, N_3, \ldots, N_{k+1}](w)$ accepts. There must be at
least one query $w'\in\sigmastar$ along $\lambda(\alpha)$ such that
for some (and hence by the previous sentence for {\em every})
$\beta\in [\alpha]$, adding $\beta$ to $\oor$ changes the answer of
$[\oor ; N_2, N_3, \ldots, N_{k+1}](w')$, since otherwise
$\lambda(\alpha)$ would also appear in $[\oor; N_1, N_2, \ldots,
N_{k+1}](x)$.
Also note that for every $A \subseteq U$ with $||A|| \leq k+1$,
$[\oor \cup A; N_2, N_3, \ldots, N_{k+1}]$ is unambiguous. Thus we
get a contradiction with the induction hypothesis, since
$||[\alpha]||> 5^{k} \cdot \prod_{i=1}^{k+1}(r \circ)^{i}(|x|) \ge
5^{k} \cdot \prod_{i=2}^{k+1}(r \circ)^{i}(|x|) \geq 5^{k} \cdot
\prod_{i=1}^{k}(r \circ)^{i}(|w'|)$. \\

\noindent {\boldmath \bf Case 2: Every equivalence class of $\sigma$
is of size $\leq 5^{k} \cdot \prod_{i=1}^{k+1}(r \circ)^{i}(|x|)$.}
We need the following claim, which demonstrates that if $\sigma$ is
an equivalence relation over some set $C$ consisting only of small
equivalence classes, then $C$ can be partitioned into two
sufficiently large disjoint sets $C_1$ and $C_2$ such that every
equivalence class is contained in either $C_1$ or $C_2$.

\begin{claim}
\label{claim:equivalence-partition} Let $\sigma$ be any equivalence
relation over some set $C$ and let $||C|| > 5s$.  If $||[\alpha]||
\le s$ for every $\alpha\in C$, then there exists a set $J\subseteq
 C$ such that $2 s < ||\bigcup_{\alpha \in J}[\alpha]|| < ||C|| - 2s$.
\end{claim}
\noindent {\bf Proof of Claim~\ref{claim:equivalence-partition}.}
Take any set $J'\subseteq C$ of maximum cardinality such that
$||\bigcup_{\alpha \in J'}[\alpha]|| \le 2s$. Let $J = J' \cup
[\beta]$, where $\beta$ is any element in $C-J'$ (note that $\beta$
exists because $||C|| > 2s$). Set $J$ witnesses the correctness of
the claim, which can be seen as follows. First, $||\bigcup_{\alpha
\in J}[\alpha]|| > 2s$ because $||J|| > ||J'||$ and $J'$ is a
maximum cardinality set such that $||\bigcup_{\alpha \in
J'}[\alpha]|| \le 2s$. Second, $||\bigcup_{\alpha \in J}[\alpha]||
\le 2s + s = 3s$ because $||\bigcup_{\alpha \in J'}[\alpha]|| \le
2s$ and $||[\beta]|| \le s$. Since $||C|| > 5s$, this implies that
$||\bigcup_{\alpha \in J}[\alpha]|| <
||C||-2s$.~\qed~(Claim~\ref{claim:equivalence-partition})

\medskip

\noindent It is easy to see that our equivalence class $\sigma$ over
$C$ satisfies the preconditions of
Claim~\ref{claim:equivalence-partition}. Hence, it follows that
there exists a set $J \subseteq C$ such that
\begin{eqnarray*}
2 \cdot 5^{k} \cdot \prod_{i=1}^{k+1}(r \circ)^{i}(|x|) <
||\bigcup_{\alpha \in J}[\alpha]|| < ||C|| - 2 \cdot 5^{k} \cdot
\prod_{i=1}^{k+1}(r \circ)^{i}(|x|).
\end{eqnarray*}

\noindent Let $C_1 := \bigcup_{\alpha \in J}[\alpha]$ and let $C_2
:= C - C_{1}$. Then both $||C_1||$ and $||C_2||$ are greater than $2
\cdot 5^{k} \cdot  \prod_{i=1}^{k+1}(r \circ)^{i}(|x|)$. For every
$\alpha_1 \in C_1$ and $\alpha_2 \in C_2$, we define
$\confl(\alpha_1) =$
\begin{eqnarray*}
\{\beta_2 \in C_2~|~\mbox{$\lambda(\alpha_1)$ is not an accepting
path in } [\oor \cup \{\alpha_1, \beta_2\}; N_1, N_2, \ldots,
N_{k+1}](x)\},
\end{eqnarray*}
and define $\confl(\alpha_2) =$
\begin{eqnarray*}
\{\beta_1 \in C_1~|~\mbox{$\lambda(\alpha_2)$ is not an accepting
path in } [\oor \cup \{\alpha_2, \beta_1\}; N_1, N_2, \ldots,
N_{k+1}](x)\}.
\end{eqnarray*}

\noindent We claim that both $||\confl(\alpha_1)||$ and
$||\confl(\alpha_2)||$ are no more than $5^{k} \cdot
\prod_{i=1}^{k+1}(r \circ)^{i}(|x|)$. To prove this, first note that
$N_1(x)$ with any oracle can ask at most $r(|x|)$ queries along a
computation path. Second, by the induction hypothesis, for every
query $w$ made by $N_1(x)$, there can be at most $5^{k} \cdot
\prod_{i=1}^{k} (r\circ)^{i}(|w|)$ $\leq$ $5^{k} \cdot
\prod_{i=2}^{k+1} (r\circ)^{i}(|x|)$ strings $\beta \in U$ such that
adding $\beta$ to some oracle $\oor'$ changes the decision
(acceptance or rejection) of $[\oor'; N_2, N_3, \ldots,
N_{k+1}](w)$. These two facts imply the stated bound on
$||\confl(\alpha_1)||$ and $||\confl(\alpha_2)||$.

With the bounds $||C_1||$, $||C_2||$ $> 2 \cdot 5^{k} \cdot
\prod_{i=1}^{k+1}(r \circ)^{i}(|x|)$, a simple counting argument now
shows that there exist $\alpha_1 \in C_1$ and $\alpha_2 \in C_2$
such that $\alpha_1 \not \in \confl(\alpha_2)$ and $\alpha_2 \not
\in \confl(\alpha_1)$. As a consequence, $[\oor \cup \{\alpha_1,
\alpha_2\}; N_1, N_2, \ldots, N_{k+1}](x)$ accepts along two
distinct paths, namely $\lambda(\alpha_1)$ and $\lambda(\alpha_2)$,
which are indeed distinct since $[\alpha_1] \neq [\alpha_2]$. This
gives a contradiction with the assumption that $[\oor \cup A; N_1,
N_2, \ldots, N_{k+1}]$ is unambiguous for every $A \subseteq U$ with
$||A|| \leq k+1$.~\qed~(Lemma~\ref{lemma:weakness-uph-machines})

\subsection{\boldmath The Notion of $(h,t)$-ambiguity for Functions on $\powerset(\sigmastar)$\unboldmath}

\noindent %
Any oracle machine can be interpreted as a function mapping a set of
strings to another set of strings as follows: A machine $N$ maps any
set $\oor$ to the set $L(N^{\oor})$. Therefore it makes sense to
consider the (possibly partial) function $\mlor :
\powerset(\sigmastar) \rightarrow \powerset(\sigmastar)$ defined by
a $\Sigma_{k}(\cdot)$-system $[\cdot; N_1, N_2, \ldots, N_k]$. (That
is, define $\mlor$ so that for every $\oor \subseteq \sigmastar$,
$\mlor(\oor) =_{df} L_{\textnormal{unambiguous}}[\oor; N_1, N_2,
\ldots, N_k]$.) We introduce a convenient notion called
``$(h,t)$-ambiguity'' for (partial or total) functions, which we
will later apply to functions defined by
$\Sigma_{k}(\cdot)$-systems.
\begin{definition}
\label{def:ht-ambiguity} For any $h \in \naturalnumber^{+}$ and
polynomial $t(\cdot)$, we call a partial or total function $\mlor :
\powerset(\sigmastar) \rightarrow \powerset(\sigmastar)$
\emph{$(h,t)$-ambiguous} if for every $\oor, U \subseteq \sigmastar$
with $\oor \cap U = \emptyset$, one of the following is true:
\begin{enumerate}
\item For some $A\subseteq U$ with $||A|| \le h$, $\mlor(\oor \cup A)$ is
  undefined, or
\item for every $w\in\sigmastar$,
\[
  ||\{ \alpha \in U \condition w \in \mlor(\oor \cup \{ \alpha \})
  \Longleftrightarrow w \notin \mlor(\oor) \}|| \le  t(|w|).
\]
\end{enumerate}
\end{definition}
The machine $N_1$ in a $\Sigma_{k}(\oor)$-system $[\oor;
N_1, N_2, \ldots, N_k]$ has oracle access to the set
$L[\oor; N_2, N_3, \ldots, N_k]$.
In many of our proofs, we first apply
Lemma~\ref{lemma:weakness-uph-machines} to prove that under certain
conditions the $\Sigma_{k-1}(\cdot)$-subsystem $[\cdot; N_2, N_3,
\ldots, N_k]$ defines a $(k,t)$-ambiguous function $\mlor'$, where
$t$ is some polynomial and for any $\oor \subseteq \sigmastar$,
$\mlor'(\oor)$ is defined to be $L[\oor; N_2, N_3, \ldots, N_k]$.
The $(k,t)$-ambiguity of the function $\mlor'$ defined by the
$\Sigma_{k-1}(\cdot)$-subsystem $[\cdot; N_2, N_3, \ldots, N_k]$ is
the only property of $\mlor'$ that is needed in our proofs.
Therefore, we can assume without loss of generality that in a
$\Sigma_{k}(\oor)$-system $[\oor; N_1, N_2, \ldots, N_k]$, machine
$N_1$ has oracle access to the set ${\mlor(\oor)}$, where $\mlor$
can be any arbitrary $(k,t)$-ambiguous function, rather than to the
set $L[\oor; N_2, N_3, \ldots, N_k]$. Using this approach has its
advantages: It greatly simplifies our proof arguments and makes
expressions compact, since we no longer need to deal with stacks of
oracle $\nptm$s.

\section{Applications}

\noindent In this section, we demonstrate applications of our proof
techniques. We show that our counting techniques are useful in
generalizing certain known relativization results involving bounded
ambiguity classes such as $\up$ and $\promiseup$ to new results
involving arbitrary levels of $\uph$. This stands in contrast to
generalizations achieved for relativization results involving levels
of $\ph$. For instance, extending the relativized separation of
initial levels of $\ph$~\cite{bak-gil-sol:j:rel,bak-sel:j:step2} to
relativized separations of arbitrary levels of
$\ph$~\cite{yao:c:separating} required applications of sophisticated
circuit-theoretic techniques. For the case of the unambiguous
polynomial hierarchy however, we show (via
Theorem~\ref{thm:up-leqk-notin-usigmapk}) that a relativized
separation of its levels can be achieved by counting techniques
alone, i.e., without resorting to circuit-theoretic tools and
techniques.

\subsection{Comparing Bounded Ambiguity Classes with the Levels of $\uph$}
\label{subsec:sep-pruph-uph}

\noindent We compare classes defined by nondeterministic
polynomial-time Turing machines having restrictions on the number of
accepting paths ($\up_{O(1)}$ and $\fewp$) with levels of the
unambiguous polynomial hierarchy ($\uph$). It is known that
$\up_{\leq k} \subseteq \usigmap{k}$ in all relativized worlds.
Theorem~\ref{thm:up-leqk-notin-usigmapk} shows the optimality of
this inclusion with respect to relativizable proof techniques.
Beigel~\cite{bei:c:up1} constructed an oracle relative to which
$\up_{k(n)+1} \not \subseteq \up_{k(n)}$, for every polynomial $k(n)
\geq 2$. Theorem~\ref{thm:up-leqk-notin-usigmapk} subsumes this
oracle result of Beigel~\cite{bei:c:up1} for any constant $k$.

By a slight modification of the oracle construction in
Theorem~\ref{thm:up-leqk-notin-usigmapk}, we can show that the
second level of the promise unambiguous hierarchy $\prusigmap{2}$
is not contained in the unambiguous polynomial hierarchy $\uph$.
Results on relativized separations of levels of some unambiguity
based hierarchy from another hierarchy have been investigated
earlier. %
Rossmanith (see~\cite{nie-ros:j:up-hierarchy}) gave a relativized
separation of $\ausigmap{k}$ from $\usigmap{k}$, for any $k \geq 2$.
Spakowski and Tripathi~\cite{spa-rah:j:unambig-alternating}
constructed an oracle relative to which $\ausigmap{k} \not \subseteq
\Pi_{k}^{p}$, for any $k \geq 1$. Our relativized separation of
$\prusigmap{2}$ from $\uph$ does not seem to be implied from these
previous results in any obvious way.

\begin{theorem}
\label{thm:up-leqk-notin-usigmapk} For any integer $k \geq 1$, there
exists an oracle $\aor$ such that $\up_{\leq k+1}^{\aor} \nsubseteq
{{\rm U}}\Sigma^{p, \aor}_{k}$.
\end{theorem}

\sproof Our test language is: $L(\aor) = \{0^n~|~\aor^{=n} \neq
\emptyset\}$. We will create an oracle $\aor$ that, for any length
$n \in \naturalnumber^+$, satisfies $||\aor^{=n}|| \leq k+1$. Let
$(N_{i,1}, N_{i,2}, \ldots, N_{i,k})$ be an enumeration of tuples,
where $N_{i,\star}$ is a nondeterministic polynomial-time oracle
Turing machine with running time bounded by some polynomial
$p(\cdot)$. Initially, $\aor := \emptyset$.

\medskip

\noindent {\boldmath \bf Stage $i$:} Choose a very large integer $n$
such that (a) $2^n > 5^k \cdot \prod_{j=1}^{k}(p \circ)^{j}(n)$, (b)
$n$ satisfies the promises made in the previous stages, (c) no
string of length $\geq n$ is ever queried in any of the previous
stages, and (d) $n$ is larger than the value in the previous stage.

If there exists a set $B \subseteq \Sigma^{n}$ such that $||B|| \leq
k+1$ and $[\aor \cup B; N_{i,1}, N_{i,2}, \ldots, N_{i,k}]$ is
not unambiguous, then set $\aor := \aor \cup B$. Promise to choose
the value of $n$ in the next stage to be larger than $(p
\circ)^k(|w|)$, where $w$ is an arbitrary string witnessing that
$[\aor \cup B; N_{i,1}, N_{i,2}, \ldots, N_{i,k}]$ is not
unambiguous, and then move to the next stage.

Otherwise, choose a string $\alpha \in \Sigma^{n}$ (as guaranteed by
Lemma~\ref{lemma:weakness-uph-machines} and by our choice of $n$)
such that
\[
[\aor; N_{i,1}, N_{i,2}, \ldots, N_{i,k}](0^n) \mbox{ accepts }
\Longleftrightarrow [\aor \cup \{\alpha\}; N_{i,1}, N_{i,2}, \ldots,
N_{i,k}](0^n) \mbox{ accepts}.
\]
If $[\aor; N_{i,1}, N_{i,2}, \ldots, N_{i,k}](0^n)$ accepts, then
move to the next stage. Otherwise, if $[\aor; N_{i,1}, N_{i,2},
\ldots, N_{i,k}](0^n)$ rejects, then set $\aor := \aor \cup
\{\alpha\}$ and move to the next stage. \\
{\bf End of Stage}

\medskip

\noindent Clearly the construction guarantees that $L(\aor) \in
\up_{\leq k+1}^{\aor} - {{\rm U}}\Sigma^{p,\aor}_{k}$.~\qed \\

\noindent A straightforward adaptation of the proof of
Theorem~\ref{thm:up-leqk-notin-usigmapk} allows to separate
the second level, $\prusigmap{2}$, of the promise unambiguous
polynomial hierarchy from the unambiguous polynomial hierarchy,
$\uph$, in some relativized world. We obtain this relativized
separation via Theorem~\ref{thm:prup-closure-notin-uph}, where the
subclass $\fewp^{\aor}$ of $\mathcal{U}\Sigma^{p,\aor}_{2}$ is
separated from $\uph^{\aor}$.

\begin{theorem}
\label{thm:prup-closure-notin-uph} There exists an oracle $\aor$
such that $\fewp^{\aor} \nsubseteq \uph^{\aor}$.
\end{theorem}

\noindent {\bf Proof Sketch:} Take the test language $L(\aor)$ used
in the proof of Theorem~\ref{thm:up-leqk-notin-usigmapk}. Maintain
the stipulation that the oracle $\aor$ satisfies, for all $n \in
\naturalnumber^+$, $||\aor^{=n}|| \leq n$; this ensures that
$L(\aor) \in \fewp^{\aor}$. Finally, for all $k \in
\naturalnumber^{+}$, diagonalize against all tuples $(N_{i,1},
N_{i,2}, \ldots, N_{i,k})$ as in the proof of
Theorem~\ref{thm:up-leqk-notin-usigmapk}.~\qed
\begin{corollary}
\label{cor:prup2-notin-uph} There is a relativized world where
$\prusigmap{2}$ is not contained in $\uph$.
\end{corollary}

\noindent Cai, Hemachandra, and
Vysko\v{c}~\cite{cai-hem-vys:b:promise} proved that smart $2$-Turing
access to $\promiseup$ cannot be subsumed by $\conp^{\up} \cup
\np^{\up}$ in some relativized world. As a consequence, they showed
that there is a relativized world where smart bounded adaptive
reductions to $\promiseup$ and smart bounded nonadaptive reductions
to $\promiseup$ are nonequivalent, a characteristic that stands in
contrast with the cases of $\up$ and $\np$. (Both $\up$ and $\np$
are known to have equivalence between bounded adaptive reductions
and bounded nonadaptive reductions in all relativized worlds
(see~\cite{cai-hem-vys:b:promise,wag:j:bounded}.) We obtain a
generalization of their result as a corollary of
Theorem~\ref{thm:up-leq-k-conup-usigmapk-1}: There is a relativized
world where smart $k$-Turing access to $\promiseup$ is not contained
in $\conp^{{{\rm U}}\Sigma^{p,A}_{k-1}} \cup \np^{{{\rm
U}}\Sigma^{p,A}_{k-1}}$, for any $k \geq 2$. The proof of
Theorem~\ref{thm:up-leq-k-conup-usigmapk-1} gives a first example
for the role of $(h,t)$-ambiguity in derivations of our results.

\begin{theorem}
\label{thm:up-leq-k-conup-usigmapk-1} For any integer $k \geq 2$,
there exists an oracle $\aor$ such that
\[
\up_{\leq k}^{\aor} \nsubseteq \conp^{{{\rm U}}\Sigma^{p,
\aor}_{k-1}}
\]
\end{theorem}

\sproof The test language $L(\aor)$ is the same as the one in the
proof of Theorem~\ref{thm:up-leqk-notin-usigmapk}. We maintain the
stipulation that for every $n \in \naturalnumber^+$, $||\aor^{=n}||
\leq k$; then, clearly $L(\aor) \in \up_{\leq k}^{\aor}$. We will
show that $L(\aor) \not \in \conp^{{{\rm U}}\Sigma^{p,A}_{k-1}}$.

Let $(N_{i,1}, N_{i,2}, \ldots, N_{i,k})$ be an enumeration of
tuples, where $N_{i,\star}$ is a nondeterministic polynomial-time
oracle Turing machine. Initially, $\aor := \emptyset$.

\medskip

\noindent {\boldmath \bf Stage $i$:} Let $p(\cdot)$ be a polynomial
that bounds the running time of $N_{i,\star}$. Choose a large
integer $n$ such that (a) $2^n > 5^{k-1} \cdot \prod_{i=1}^{k} (p
\circ)^i(n)$, (b) $n$ satisfies any promises made in the previous
stages, (c) $n$ is larger than the value of $n$ in the previous
stages, and (d) no queries of length $\geq n$ are made in the
previous stages.

If there exists a set $B \subseteq \Sigma^n$ such that $||B||
\leq k$ and $[\aor \cup B; N_{i,2}, N_{i,3}, \ldots, N_{i,k}]$
is not unambiguous, then set $\aor := \aor \cup B$. Promise to
choose the value of $n$ in the next stage to be larger than
$(p\circ)^{k-1}(|w|)$, where $w$ is any arbitrary string witnessing
that $[\aor \cup B; N_{i,2}, N_{i,3}, \ldots, N_{i,k}]$ is not
unambiguous, and then move to the next stage.

Otherwise proceed as follows. Let $\mlor: \powerset(\sigmastar)
\rightarrow \powerset(\sigmastar)$ be the function defined so that
for every $\oor \subseteq\sigmastar$,
\[
  \mlor(\oor) =_{df} L_{\textnormal{unambiguous}}[\oor ; N_{i,2}, N_{i,3}, \ldots , N_{i,k}].
\]
It follows from Lemma~\ref{lemma:weakness-uph-machines} that $\mlor$ is
$(k,t)$-ambiguous for some polynomial $t(\cdot) =_{df}
5^{k-1} \cdot \prod_{i=1}^{k-1} (p \circ)^{i}(\cdot)$. Note that
$\mlor(\aor\cup B)$ is defined for every $B\subseteq \Sigma^n$
satisfying $||B|| \le k$.

If $N_{i,1}^{\mlor(\aor)}(0^n)$ rejects, then move to the next
stage. Otherwise, fix an accepting path $\rho$ in
$N_{i,1}^{\mlor(\aor)}(0^n)$. For each query $w \in Q(\rho)$, let
$C(w) =_{df} \{\alpha \in \Sigma^n~|~w \in \mlor(\aor \cup
\{\alpha\}) \Longleftrightarrow w \not \in \mlor(\aor)\}$. By the
definition of $(k,t)$-ambiguity, for each $w \in Q(\rho)$, we have
$||C(w)|| \leq t(|w|) \leq t(p(n))$. Because $n$ is large enough, we
can choose some $\alpha \in \Sigma^n$ such that $\alpha \not \in
\bigcup_{w \in Q(\rho)} C(w)$. Set $\aor := \aor \cup
\{\alpha\}$ and move to the next stage. \\
{\bf End of Stage}

\medskip

\noindent Clearly the construction guarantees that $L(\aor) \not \in \conp^{{{\rm
U}}\Sigma^{p,A}_{k-1}}$.~\qed

\begin{corollary}
\label{cor:k-turing-promiseup} For any integer $k \geq 2$, there
exists an oracle $\aor$ such that
\[
R^{p}_{s,k\hbox{-}T}(\promiseup^{\aor}) \nsubseteq \conp^{{{\rm
U}}\Sigma^{p,A}_{k-1}} \cup \np^{{{\rm U}}\Sigma^{p,A}_{k-1}}.
\]
\end{corollary}
\textbf{Proof} This follows from
Theorem~\ref{thm:up-leq-k-conup-usigmapk-1} because for all oracles
$\aor$, $\up_{\leq k}^{\aor} \subseteq
R^{p}_{s,k\hbox{-}T}(\promiseup^{\aor})$ and
$R^{p}_{s,k\hbox{-}T}(\promiseup^{\aor})$ is closed under
complementation.~\qed

\subsection{Simulating Nonadaptive Access by Adaptive Access (Non-promise Case)}
\label{subsec:nonadap-adap-nonpromise}

\noindent It is known that adaptive Turing access to $\np$ is
exponentially more powerful compared to nonadaptive Turing access to
$\np$. That is, $R^{p}_{(2^k-1)\hbox{-}tt}(\np) \subseteq
R^{p}_{k\hbox{-}T}(\np)$~\cite{bei:j:bounded-queries} and this
inclusion relativizes. However, for the case of unambiguous
nondeterministic computation such a relationship between nonadaptive
access and adaptive access is not known. Cai, Hemachandra, and
Vysko\v{c}~\cite{cai-hem-vys:c-OUT-BY-BOOK:promise} showed that even
proving the superiority of adaptive Turing access over nonadaptive
Turing access with one single query more might be nontrivial for
unambiguous nondeterministic computation:
\begin{theorem}[\cite{cai-hem-vys:c-OUT-BY-BOOK:promise}]
\label{thm:CHV1} For any total, polynomial-time
computable and polynomially bounded function $k(\cdot)$,
there exists an oracle $\aor$ such that
\[
R^{p}_{(k(n)+1)\hbox{-}tt}(\up^\aor)
\not \subseteq R^{p, \aor}_{k(n)\hbox{-}T}(\up^\aor).
\]
\end{theorem}
In the next theorem, we generalize this result to the higher levels of the
unambiguous polynomial hierarchy $\uph$.

\begin{theorem}
\label{thm:truth-table-versus-Turing} For any total, polynomial-time
computable and polynomially bounded function $k(\cdot)$, and integer
$h \geq 1$, there exists an oracle $\aor$ such that
\[
\red{(k(n)+1)\hbox{-}dtt}{p}{\up_{\le h}^\aor } \not\subseteq
\red{k(n)\hbox{-}T}{p, \aor}{ {\rm U}\Sigma^{p,\aor}_{h} },
\]
and hence $\red{(k(n)+1)\hbox{-}dtt}{p}{  {\rm
U}\Sigma^{p,\aor}_{h}} \not\subseteq \red{k(n)\hbox{-}T}{p, \aor}{
{\rm U}\Sigma^{p,\aor}_{h}      }$.
\end{theorem}

\noindent
For the proof of Theorem~\ref{thm:truth-table-versus-Turing}, we need the following lemma.

\begin{lemma}
\label{lemma:sensitive-htambiguity} Fix an oracle $\nptm$ $N$ with
running time bounded by some polynomial $p(\cdot)$, string $x \in
\sigmastar$, and sets $\oor, U_1, U_2 \subseteq \sigmastar$ such
that $\oor\cap U_1 = \oor\cap U_2 =  U_1 \cap U_2 = \emptyset$. Let
$\mlor : \powerset(\sigmastar) \rightarrow \powerset(\sigmastar)$ be
an $(h,t)$-ambiguous function such that $\mlor(\oor \cup A_1 \cup
A_2)$ is defined for every $A_1 \subseteq U_1$ and $A_2 \subseteq
U_2$ with $||A_1|| \le h$ and $||A_2|| \leq h$. Let
\begin{eqnarray*}
  C_1 & = & \{ \alpha\in U_1 \condition
       N^{\mlor(\oor) }(x)  \text{ accepts} \Longleftrightarrow
       N^{\mlor(\oor \cup \{ \alpha \}) }(x)
              \text{ rejects} \} \mbox{ and}\\
  C_2 & = & \{ \alpha\in U_2 \condition
       N^{\mlor(\oor) }(x)  \text{ accepts} \Longleftrightarrow
       N^{\mlor(\oor \cup \{ \alpha \}) }(x)
              \text{ rejects} \}.
\end{eqnarray*}
If $N^{\mlor(\oor \cup A_1 \cup A_2)}(x)$ is unambiguous for every
$A_1 \subseteq U_1$ and $A_2 \subseteq U_2$ with $||A_1|| \leq 1$
and $||A_2|| \leq 1$, then $\min\{||C_1||, ||C_2||\} \leq 2 \cdot
p(|x|)\cdot t(p(|x|)) \cdot (p(|x|)\cdot t(p(|x|)) + 1)$.
\end{lemma}
\sproof
We start with the easier case that $N^{\mlor(\oor)}(x)$ accepts. Let
$\rho$ be the (unique) accepting computation path in
$N^{\mlor(\oor)}(x)$.
Then $N^{\mlor(\oor \cup \{\alpha\}) }(x)$ accepts unless for some
query $w \in Q(\rho)$, it is the case that $w \in \mlor(\oor \cup \{
\alpha \}) \Longleftrightarrow w \notin \mlor(\oor)$. Since
$N^{\mlor(\oor)}(x)$ queries at most $p(|x|)$ queries along every
path, since each query $w \in Q(\rho)$ is of length $\leq p(|x|)$,
since $\mlor$ is $(h,t)$-ambiguous, and since $\mlor(\oor \cup A_1
\cup A_2)$ is defined for every $A_1 \subseteq U_1$ and $A_2
\subseteq U_2$ with $||A_1||$, $||A_2|| \leq h$, there cannot be
more than $p(|x|) \cdot t(p(|x|))$ strings $\alpha \in U_1$ (or,
$\alpha \in U_2$) making $N^{\mlor(\oor \cup \{\alpha\}) }(x)$
reject. Hence $||C_1|| \le p(|x|)\cdot t(p(|x|))$ and $||C_2||\le
p(|x|)\cdot t(p(|x|))$.

For the other case, suppose that $N^{\mlor(\oor)}(x)$ rejects. To get
a contradiction, assume that $\min\{||C_1||, ||C_2||\} >
2 \cdot p(|x|)\cdot t(p(|x|)) \cdot (p(|x|)\cdot t(p(|x|)) + 1)$.
For every $\alpha\in C_1 \cup C_2$, denote by $s(\alpha)$ the unique
accepting computation path in $N^{\mlor(\oor\cup \{ \alpha\})}(x)$.
Define an equivalence relation $\sigma_1$ on $C_1$ as follows: For
all $\alpha_1, \alpha_2 \in C_1$,
\[
   \alpha_1 \sigma_1 \alpha_2 \Longleftrightarrow s(\alpha_1) =
   s(\alpha_2).
\]
Let $C_1/\sigma_1$ be the quotient set of $C_1$ determined by $\sigma_1$.
We first prove that
\begin{equation} \label{many-equivalence-classes}
   ||C_1/\sigma_1 || \ge \frac{||C_1||}{p(|x|)\cdot t(p(|x|))}.
\end{equation}
To this end, consider any $\alpha \in C_1$. Note that for every
query $w\in Q(s(\alpha))$, there cannot be more than $t(|w|)$
different strings $\alpha' \in C_1$ such that
\[
 w \in \mlor(\oor \cup \{ \alpha' \})
  \Longleftrightarrow w \notin \mlor(\oor ).
\]
This holds because $\mlor$ is $(h,t)$-ambiguous and because
$\mlor(\oor \cup A)$ is defined for every $A \subseteq U_1$ with
$||A|| \leq h$. Also since there are at most $p(|x|)$ queries $w \in
Q(s(\alpha))$ and each such query is of length $\leq p(|x|)$,
there cannot be more than $p(|x|) \cdot t(p(|x|))$ strings $\alpha'
\in C_1$ such that for some query $w \in Q(s(\alpha))$, it is the
case that $w \in \mlor(\oor \cup \{ \alpha' \})$ if and only if $w
\notin \mlor(\oor)$. In other words, for all but $p(|x|) \cdot
t(p(|x|))$ strings $\alpha' \in C_1$, the membership in
$\mlor(\oor)$ of every query $w \in Q(s(\alpha))$ remains unchanged
on inclusion of $\alpha'$ to $\oor$. Since by assumption,
$N^{\mlor(\oor)}(x)$ rejects, it follows that for no more than
$p(|x|) \cdot t(p(|x|))$ strings $\alpha' \in C_1$, (accepting) path
$s(\alpha)$ appears in $N^{\mlor(\oor \cup \alpha')}(x)$. Hence
there cannot be more than $p(|x|) \cdot t(p(|x|))$ different strings
$\alpha'\in C_1$ such that $s(\alpha') = s(\alpha)$. Thus we have
proved Statement~(\ref{many-equivalence-classes}).

Analogously, the same can be proved for $C_2$ with appropriately
defined equivalence class $\sigma_2$.

Define  $\widetilde{C}_1$ to be a maximal subset of $C_1$ such that
$s(\alpha_1) \not= s(\alpha_2)$ for every $\alpha_1, \alpha_2\in
\widetilde{C}_1$ with $\alpha_1 \not= \alpha_2$.
Analogously, define $\widetilde{C}_2$.
Clearly,
\begin{equation}
\label{eq:lower-bound-tilde-C1}
  ||\widetilde{C}_1 || \ge \frac{||C_1||}{p(|x|)\cdot t(p(|x|))}
  >
  2 \cdot (p(|x|)\cdot t(p(|x|)) + 1),
\end{equation}
and
\begin{equation}
\label{eq:lower-bound-tilde-C2}
  ||\widetilde{C}_2 || \ge \frac{||C_2||}{p(|x|)\cdot t(p(|x|))}
  >
  2 \cdot (p(|x|)\cdot t(p(|x|)) + 1).
\end{equation}
For every $\alpha_1 \in \widetilde{C}_1$, let
\[
\confl(\alpha_1) = \{\beta_2 \in \widetilde{C}_2 \condition
s(\alpha_1) \textnormal{ does not appear in } N^{\mlor(\oor\cup \{
\alpha_1, \beta_2\})}(x) \mbox{ or $s(\alpha_1) = s(\beta_2)$}\},
\]
and for every $\alpha_2 \in \widetilde{C}_2$, let
\[
\confl(\alpha_2) = \{\beta_1 \in \widetilde{C}_1 \condition
s(\alpha_2) \textnormal{ does not appear in } N^{\mlor(\oor\cup \{
\alpha_2, \beta_1\})}(x) \mbox{ or $s(\alpha_2) = s(\beta_1)$}\}.
\]
Fix an arbitrary $\alpha \in \widetilde{C}_1$. Since $\mlor$ is
$(h,t)$-ambiguous and $\mlor(\oor \cup A_1 \cup A_2)$ is defined for
every $||A_1|| \leq 1$ and $||A_2|| \leq h$, it holds that for every
query $w \in Q(s(\alpha))$ in $N^{\mlor(\oor \cup \{\alpha\})}(x)$,
there cannot be more than $t(|w|)$ different strings $\beta \in
\widetilde{C}_2$ such that
\[
w \in \mlor(\oor \cup \{\alpha, \beta\}) \Longleftrightarrow w
\not \in \mlor(\oor \cup \{\alpha\}).
\]
Also, since no more than $p(|x|)$ strings are queried on each path
in $N^{\mlor(\oor \cup \{\alpha\})}(x)$, since each queried string
is of length $\leq p(|x|)$, and since for distinct $\alpha_1$ and
$\alpha_2$, $s(\alpha_1) \neq s(\alpha_2)$ whenever $\{\alpha_1,
\alpha_2\} \subseteq \widetilde{C}_2$, we have that for any
$\alpha\in \widetilde{C}_1$,
\[
||\confl(\alpha)|| \leq p(|x|)\cdot t(p(|x|)) + 1.
\]
Analogously, we can prove that for any $\alpha' \in
\widetilde{C}_2$,
\[
||\confl(\alpha')|| \leq p(|x|)\cdot t(p(|x|)) + 1.
\]
With the lower bounds on $||\widetilde{C}_1||$ and
$||\widetilde{C}_2||$ given by Eq.~(\ref{eq:lower-bound-tilde-C1})
and (\ref{eq:lower-bound-tilde-C2}), it now follows by a simple
counting argument that there is a pair $(\alpha_1, \alpha_2) \in
\widetilde{C}_1 \times \widetilde{C}_2$ such that $\alpha_2\notin
\confl(\alpha_1)$ and $\alpha_1\notin \confl(\alpha_2)$. Take two
such strings $\alpha_1$ and $\alpha_2$. It is easy to see that both
$s(\alpha_1)$ and $s(\alpha_2)$ appear in $N^{\mlor(\oor\cup \{
\alpha_1, \alpha_2\})}(x)$. Furthermore, $s(\alpha_1) \not=
s(\alpha_2)$. Hence, $N^{\mlor(\oor\cup \{ \alpha_1,
\alpha_2\})}(x)$ has two different accepting computation paths:
$s(\alpha_1)$ and $s(\alpha_2)$. This gives a contradiction to the
assumption that $N^{\mlor(\oor \cup A_1 \cup A_2)}$ is unambiguous
whenever $A_1 \subseteq U_1$ and $A_2 \subseteq U_2$ with $||A_1||
\leq 1$ and $||A_2|| \leq
1$.~\qed(Lemma~\ref{lemma:sensitive-htambiguity})

\bigskip

\noindent {\bf Proof of
Theorem~\ref{thm:truth-table-versus-Turing}.\quad} The test language
$L(\aor)$ for our oracle construction is inspired by the one
in~\cite[Theorem 3.1]{cai-hem-vys:b:promise}. For length $n$, we
will reserve the following segment of $k(n)+1$ regions $S_{n,f} =
1^n01^f0\Sigma^n$, where $f\in [k(n)+1]$. For $n\ge 1$, define $S_n
= \bigcup_{f=1}^{k(n)+1} S_{n,f}$. For all $n\ge 1$ and $f\in
[k(n)+1]$, we stipulate that $||\aor\cap S_{n,f}|| \le h$. Let
\[
  L(\aor) = \{ 0^n \condition ||\aor\cap S_n|| \ge 1\}.
\]
Clearly, as long as the oracle set $\aor$ maintains the stipulation
that $||\aor\cap S_{n,f}|| \le h$, we have $L(\aor) \in
\red{(k(n)+1)\hbox{-}dtt}{p}{
  \up_{\le h}^\aor }$
and hence $L(\aor) \in\red{(k(n)+1)\hbox{-}dtt}{p}{  {\rm
U}\Sigma^{p,\aor}_{h}  }$. We construct an oracle $\aor$ such that
$L(\aor) \notin \red{k(n)\hbox{-}T}{p}{ {\rm U}\Sigma^{p,\aor}_{h} }$.
\footnote{The construction can be easily modified to prove the stronger result
that $L(\aor) \notin \red{k(n)-T}{p, \aor}{{\rm
U}\Sigma^{p,\aor}_{h}}$.}

We give a brief informal outline of how the diagonalization for
$L(\aor) \notin \red{k(n)\hbox{-}T}{p}{ {\rm U}\Sigma^{p,\aor}_{h}
}$ is carried out. Fix some input $0^n$. The crucial fact is that we
have $k(n)+1$ regions but only $k(n)$ adaptive Turing queries. There
are two cases. The easier case is when we can destroy the
unambiguity of one of the machines defining the ${\rm
U}\Sigma^{p,\aor}_{h}$ set by adding some strings to the current
segment (but, of course without violating the above stipulation,
i.e., the stipulation that $||\aor \cap S_{n,f}|| \leq h$ for each
$f \in [k(n)+1]$). In that case we can simply add these strings to
the oracle and move to the next stage. Otherwise, we can use
Lemma~\ref{lemma:sensitive-htambiguity} to show that each Turing
query is \emph{insensitive} to all but one of the $k(n)+1$ regions.
A Turing query $\beta$ is {\em insensitive} to a region if adding a
single string $\alpha$ to that region does not change the answer to
$\beta$, unless the string $\alpha$ comes from a very small
(i.e.,~polynomially bounded) number of exceptions. But we have only
$k(n)$ Turing queries. That's why, there must be a region $S_{n,f}$
to that all Turing queries are insensitive. Since $S_{n,f}$ has
exponentially many strings but only polynomially many exceptions,
there must be at least one string $\alpha^* \in S_{n,f}$ that we can
add to the current segment without changing the answers to any of
the Turing queries, and hence without changing the decision (i.e,
acceptance or rejection) of the deterministic machine $M$ making the
Turing reduction. We add this string $\alpha^*$ to the oracle
(thereby changing the membership of $0^n$ in the test language) if
and only if $M$ rejects. This completes the construction in the
current stage, and so we move to the next stage.

Now we come to the formal description of the diagonalization.
Let $(N_{i,1},N_{i,2},\ldots , N_{i,h} , M_j)$ be an enumeration of
tuples where $N_{i, *}$ is a nondeterministic polynomial-time oracle
Turing machine, and $M_j$ is a deterministic polynomial-time oracle
Turing machine making, for any set $\aor$ and any input of length
$n$, at most $k(n)$ queries to $L[\aor; N_{i,1}, N_{i,2}, \ldots ,
N_{i,h}]$ and at most polynomially many queries to $\aor$.
Initially, let $\aor:=\emptyset$.

\medskip

\noindent {\bf Stage \boldmath $\langle i, j\rangle$:\unboldmath }
Let $p(\cdot)$ be a polynomial that bounds the running time of both
$N_{i,*}$ and $M_j$. Choose an integer $n$ satisfying the following
requirements: (a) $2^n > 2 \cdot k(n) \cdot p(p(n)) \cdot t(p(p(n)))
\cdot (p(p(n)) \cdot t(p(p(n))) + 1)$, where $t(\cdot)$ is a
polynomial defined later in this proof, (b) $n$ is large enough so
that $n$ satisfies any promises made in the previous stages and no
string of length greater than or equal to $n$ is queried in any of
the previous stages, and (c) $n$ is larger than the value of $n$ in
the previous stage.

If there exists a set $B\subseteq S_n$ satisfying $||B\cap S_{n,f}||
\le h$ for every $f\in [k(n)+1]$ such that $[\aor\cup B; N_{i,1},
N_{i,2}, \ldots , N_{i,h}]$ is not unambiguous, then set $\aor :=
\aor \cup B$. Promise to choose the value of $n$ in the next stage
to be larger than $(p \circ)^h(|w|)$, where $w$ is an arbitrary
input string witnessing that $[\aor\cup B; N_{i,1}, N_{i,2}, \ldots
, N_{i,h}]$ is not unambiguous, and then move to the next stage.

Otherwise proceed as follows. Let $\mlor: \powerset(\sigmastar)
\rightarrow \powerset(\sigmastar)$ be the function defined so that
for every $\oor \subseteq\sigmastar$,
\[
  \mlor(\oor) =_{df} L_{\textnormal{unambiguous}}[\oor ; N_{i,2}, N_{i,3}, \ldots , N_{i,h}].
\]
By Lemma~\ref{lemma:weakness-uph-machines}, $\mlor$ is
$(h,t)$-ambiguous for some polynomial $t(\cdot)$. (To be specific,
$t(\cdot) = 5^{h-1} \cdot \prod_{i=1}^{h-1} (p \circ)^{i}(\cdot)$.)
Note that $\mlor(\aor\cup B)$ is defined for every $B\subseteq S_n$
satisfying $||B\cap S_{n,f}|| \le h$ for every $f\in [k(n)+1]$.

If $M_j(0^n)$ with oracle $L(N_{i,1}^{\mlor(\aor)})$ is accepting,
then move on to the next stage. If $M_j(0^n)$ with oracle
$L(N_{i,1}^{\mlor(\aor)})$ is rejecting, then look for a string
$\alpha \in S_n$ that can be added to $\aor$ without changing the
decision (i.e., acceptance or rejection) of $M_j(0^n)$ with oracle
$L(N_{i,1}^{\mlor(\aor)})$. Set $\aor:= \aor \cup \{ \alpha \}$ and
move
to the next stage.\\
{\bf End of Stage}

\medskip

\noindent It remains to show that such a string $\alpha$ always
exists. Consider $M_j(0^n)$ with oracle $L(N_{i,1}^{\mlor(\aor)})$.
Let $\beta_1, \beta_2, \ldots , \beta_{k(n)}$ be the sequence of
queries made by $M_j(0^n)$ to the oracle $L(N_{i,1}^{\mlor(\aor)})$.
The following claim states that, for any $\beta \in \sigmastar$,
there is one special region $S_{n,\sens(\beta)}$ such that, for all
regions $S_{n,f}$ different from $S_{n,\sens(\beta)}$, and for all
but polynomially many $\alpha \in S_{n,f}$, the decision (i.e,
acceptance or rejection) of $N_{i,1}^{\mlor(\aor)}(\beta)$ remains
unchanged on addition of $\alpha$ to $\aor$.
\begin{claim}
\label{claim:sensitive-htambiguity} For each $\beta\in\sigmastar$,
there is an integer $\sens(\beta) \in [k(n)+1]$ such that the
following is true:

For every $f \in [k(n)+1] - \{ \sens(\beta) \}$, there is a set
$C_f(\beta)\subseteq S_{n, f}$ with $||C_f(\beta)|| \leq 2 \cdot
p(|\beta|)\cdot t(p(|\beta|)) \cdot (p(|\beta|)\cdot t(p(|\beta|)) +
1)$ such that, for every $\alpha\in S_{n,f} - C_f(\beta)$,
\[
   \beta \in L(N_{i,1}^{\mlor(\aor)}) \Longleftrightarrow \beta \in L(N_{i,1}^{\mlor(\aor\cup \{
   \alpha \}) }).
\]
\end{claim}
Let us assume that the claim is true. There are $k(n)+1$ regions in
$S_n$, but only $k(n)$ queries $\beta_1, \beta_2, \ldots ,
\beta_{k(n)}$ made by $M_j(0^n)$ to $L(N_{i,1}^{\mlor(\aor)})$. Let
$\ell\in [k(n)+1]$ such that $\ell \not= \sens(\beta_e)$ for every
$e\in [k(n)]$. Let $C = C_\ell(\beta_1) \cup C_\ell(\beta_2) \cup
\cdots \cup C_\ell(\beta_k)$. It is easy to see that we can add any
string in $S_{n, \ell} - C$ to $\aor$ without changing the decision
of $N_{i,1}^{\mlor(\aor)}(\beta_e)$ for any $e\in [k(n)]$.

Let $\alpha$ be any string in $S_{n, \ell} - C$. Such a string
exists because $2 \cdot k(n) \cdot p(|\beta|)\cdot t(p(|\beta|))
\cdot (p(|\beta|)\cdot t(p(|\beta|)) + 1) < 2^n$.
Clearly, $M_j(0^n)$ with its $k(n)$ queries to
$L(N_{i,1}^{\mlor(\aor)})$ accepts if and only if $M_j(0^n)$ with
its $k(n)$ queries to $L(N_{i,1}^{\mlor(\aor\cup \{ \alpha \})})$
accepts. \qed(Theorem~\ref{thm:truth-table-versus-Turing})

\bigskip

\noindent {\bf Proof of Claim~\ref{claim:sensitive-htambiguity}.}
\quad Suppose the claim were false. Then there exists $\beta \in
\sigmastar$ and $f_1, f_2 \in [k(n)+1]$ with $f_1 \not= f_2$ such
that
\begin{itemize}
\item
$
C_{f_1}(\beta) =_{df} ||\{ \alpha \in S_{n,f_1} \condition \beta \in L(N_{i,1}^{\mlor(\aor)}) \Longleftrightarrow \beta \notin L(N_{i,1}^{\mlor(\aor\cup \{
   \alpha \}) }) \}|| > 2 \cdot p(|\beta|)\cdot t(p(|\beta|)) \cdot (p(|\beta|)\cdot t(p(|\beta|)) + 1),
$ and
\item
$
  C_{f_2}(\beta) =_{df} ||\{ \alpha \in S_{n,f_2} \condition \beta \in L(N_{i,1}^{\mlor(\aor)}) \Longleftrightarrow \beta \notin L(N_{i,1}^{\mlor(\aor\cup \{
   \alpha \}) }) \}|| > 2 \cdot p(|\beta|)\cdot t(p(|\beta|)) \cdot (p(|\beta|)\cdot t(p(|\beta|)) + 1).
$
\end{itemize}
Apply Lemma~\ref{lemma:sensitive-htambiguity} with $\oor := \aor$,
$U_1 := S_{n,f_1}$, and $U_2 := S_{n,f_2}$. We obtain
$\min\{||C_{f_1}(\beta)||, ||C_{f_2}(\beta)||\} \leq 2 \cdot p(|\beta|)\cdot t(p(|\beta|)) \cdot
(p(|\beta|)\cdot t(p(|\beta|)) + 1)$,
a contradiction.~\qed(Claim~\ref{claim:sensitive-htambiguity})

\subsection{Simulating Nonadaptive Access by Adaptive Access (Promise Case)}
\label{subsec:nonadap-adap-promise}

\noindent Cai, Hemachandra, and Vysko\v{c}~\cite{cai-hem-vys:b:promise}
proved the following partial improvement of their Theorem~\ref{thm:CHV1}.
\begin{theorem}[\cite{cai-hem-vys:b:promise}]
\label{thm:CHV2} For any constant $k$,
there exists an oracle $\aor$ such that
\[
R^{p}_{(k+1)\hbox{-}tt}(\up^\aor)
\not \subseteq R^{p, \aor}_{s,k\hbox{-}T}(\promiseup^\aor).
\]
\end{theorem}
Note that we have replaced ``$\up$'' by ``$\promiseup$'' on the
righthand side of the noninclusion relation of
Theorem~\ref{thm:CHV1}. This is a significant improvement for the
following reason. The computational powers of $R^{p}_{b}(\up)$ and
$R^{p}_{s,b}(\promiseup)$ (the bounded Turing closure of $\up$ and
the bounded smart Turing closure of $\promiseup$, respectively) are
known to be remarkably different in certain relativized worlds.
While it is easy to show that $\up_{\leq k}$ is robustly (i.e., for
every oracle) contained in $\p_{s,k\hbox{-}T}^{\promiseup}$ for any
$k \geq 1$, we have shown in the proof of
Theorem~\ref{thm:up-leq-k-conup-usigmapk-1} that for no $k \geq 2$,
$\up_{\leq k}$ is robustly contained in $\p^{\up}$. Therefore, it is
not immediately clear whether this improvement is impossible,
i.e.,~whether $R^{p}_{(k+1)\hbox{-}tt}(\up) \subseteq
R^{p}_{s,k\hbox{-}T}(\promiseup)$ holds relative to all oracles.

However, Cai, Hemachandra, and
Vysko\v{c}~\cite[Theorem~3.1]{cai-hem-vys:b:promise} could achieve
this improvement only by paying a heavy price. In their own words:
\begin{quote}
In our earlier version dealing with $\up^{A}$, the constant $k$ can
be replaced by any arbitrary polynomial-time computable function
$f(n)$ with polynomially bounded value. It remains open whether the
claim of the current strong version of Theorem 3.1 can be similarly
generalized to non-constant access.
\end{quote}
We resolve this open question. We show that Theorem~\ref{thm:CHV2} holds with
constant $k$ replaced by any total, polynomial-time computable and polynomially
bounded function $k(\cdot)$. This result is subsumed as the special case $h=1$ of our
main result, Theorem~\ref{thm:promis-truth-table-versus-Turing}, of this subsection.
\begin{theorem}
\label{thm:promis-truth-table-versus-Turing} For any total,
polynomial-time computable and polynomially bounded function
$k(\cdot )$, and integer $h \geq 1$, there exists an oracle $\aor$
such that
\[
\red{(k(n)+1)\hbox{-}dtt}{p}{\up_{\leq h}^{\aor}} \not\subseteq
\red{s,k(n)\hbox{-}T}{p, \aor}{\promiseup^{{\rm
U}\Sigma^{p,\aor}_{h-1}}},
\]
and hence
  $\red{(k(n)+1)\hbox{-}dtt}{p}{  {\rm U}\Sigma^{p,\aor}_{h}  } \not\subseteq
  \red{s,k(n)\hbox{-}T}{p, \aor}{ {\rm \promiseup}^{{{\rm U}}\Sigma^{p,\aor}_{h-1}}      }$.
\end{theorem}
Theorem~\ref{thm:promis-truth-table-versus-Turing} is furthermore a generalization of
Theorem~\ref{thm:CHV2} to higher levels of the unambiguous polynomial hierarchy.

The proof of Theorem~\ref{thm:promis-truth-table-versus-Turing} is
much more challenging than the proof of
Theorem~\ref{thm:truth-table-versus-Turing} because we now require
diagonalizing against $\red{s,k(n)\hbox{-}T}{p,
\aor}{\promiseup^{{\rm U}\Sigma^{p,\aor}_{h-1}}}$ as opposed to
diagonalizing against $R^{p, \aor}_{k(n)\hbox{-}T}({\up^{{\rm
U}\Sigma^{p,\aor}_{h-1}}})$. To diagonalize against $R^{p,
\aor}_{k(n)\hbox{-}T}({\up^{{\rm U}\Sigma^{p,\aor}_{h-1}}})$ as in
the proof of Theorem~\ref{thm:truth-table-versus-Turing}, it was
sufficient at any stage to extend the current oracle $\aor$ so that
the $\Sigma_{h}(\aor)$-system (corresponding to the stage) becomes
ambiguous on some input string, even if the input string witnessing
the ambiguity of the $\Sigma_{h}(\aor)$-system would never arise in
a valid computation of the deterministic querying machine
(corresponding to the same stage). This is, however, not sufficient
when we diagonalize against $\red{s,k(n)\hbox{-}T}{p,
\aor}{\promiseup^{{\rm U}\Sigma^{p,\aor}_{h-1}}}$. Any input string
witnessing the ambiguity of the $\Sigma_{h}(\aor)$-system must now
have its origin from a valid computation of the deterministic
querying machine.

To prove Theorem~\ref{thm:CHV2}, Cai et al.~\cite{cai-hem-vys:b:promise}
presented the following combinatorial lemma.
\begin{lemma}[The Gaming Lemma~\cite{cai-hem-vys:b:promise}]
\label{lemma:gaming-lemma} For $1 \leq i \leq m$, let
$\mathcal{S}_i$ be a collection of nonempty subsets of $[n]$ with
the following properties:
\begin{enumerate}
\item $(\forall j \in [n])(\exists \ell \in [m])[\{j\} \in
\mathcal{S}_{\ell}]$, and
\item $(\forall j \in [m])\left[(A, B \in S_{j} \text{ and } A \neq B) \Longrightarrow (\exists
\ell \in [j-1])(\exists C\in\mathcal{S}_{\ell}) [C \subseteq A \cup
B]\right]$.
\end{enumerate}
Then $m \geq n$.
\end{lemma}
Cai, Hemachandra, and Vysko\v{c}~\cite{cai-hem-vys:b:promise} gave
the following informal interpretation of this lemma. Suppose a
combinatorial game is to be played given the set $[n]$ at hand. The
game is played in steps by a single player. At each step $i$, the
player can generate a collection $\mathcal{S}_i$ of nonempty subsets
of $[n]$ with a restriction: If sets $A, B \subseteq [n]$ are
generated at some step $i
> 1$, then there must be a previous step $j< i$ and a set $C$
generated at that step such that $C \subseteq A \cup B$. The game
ends as soon as all the singletons $\{k\} \subseteq [n]$ are
eventually produced. The gaming lemma states that this combinatorial
game requires at least $n$ steps. %

Our proof of Theorem~\ref{thm:promis-truth-table-versus-Turing} also
makes use of the gaming lemma.
However, the actual diagonalization steps are considerably different
from the one in~\cite{cai-hem-vys:b:promise}. The most tricky part
is the proof of Lemma~\ref{lemma:existential-strings}. It
demonstrates the existence of especially \emph{nice} strings
$\alpha_1, \alpha_2, \ldots, \alpha_{r}$ satisfying certain useful
properties. Each string $\alpha_i$ serves as a representative for
one region $S_{n,i}$ of the oracle. These strings satisfy a kind of
\emph{independence} property in the following sense. Let $A$ and $B$
be two different minimal subsets of $\{\alpha_1, \alpha_2, \ldots,
\alpha_r\}$ that are minimal in the sense that adding $A$ or $B$ to
an oracle makes an oracle $\nptm$ $N$ to accept, but adding any
proper subset of $A$ or $B$ to the oracle makes $N$ to reject. Then
the independence property implies that adding all the strings in
$A\cup B$ to the oracle will make $N$ to have at least two accepting
paths.

In each stage of the diagonalization, there are two cases. The
easier case is when we can destroy the unambiguity of the ${{\rm
U}}\Sigma^{p,\aor}_{h-1}$ oracle in the $\np^{{{\rm
U}}\Sigma^{p,\aor}_{h-1}}$-machine, the machine against that we are
diagonalizaing, by adding some strings (but, of course without
violating certain requirements of the stage) to the oracle $\aor$.
In that case, we can simply add these strings to the oracle $\aor$
and move to the next stage. Otherwise, we apply the
gaming lemma (Lemma~\ref{lemma:gaming-lemma}) %
to show that by adding only the \emph{nice} strings
$\alpha_i$ to the oracle $\aor$, the desired diagonalization step
for the current stage can be achieved.
The existence of these \emph{nice} strings $\alpha_i$ was a key idea that led us to the
resolution of the question by Cai, Hemachandra, and
Vysko\v{c}~\cite{cai-hem-vys:b:promise} and in generalizing their
result.

\begin{lemma}
\label{lemma:existential-strings} Fix an oracle $\nptm$ $N$ with
running time bounded by some polynomial $p(\cdot)$, a set
$\mathcal{O} \subseteq \sigmastar$, and sets $X = \{x_1, x_2,
\ldots, x_d\}$ $\subseteq (\Sigma^*)^{\leq m}$ and $Y = \{y_1, y_2,
\ldots, y_{d'}\}$ $\subseteq (\Sigma^*)^{\leq m}$. Let $U_1, U_2,
\ldots, U_{r} \subseteq \sigmastar$ be such that for each distinct
$U_{\ell}$, $U_{\ell'}$, it holds that $\oor \cap U_{\ell}$ = $\oor
\cap U_{\ell'}$ = $U_{\ell} \cap U_{\ell'}$ = $\emptyset$ and
$||U_{\ell}||$ $=$ $||U_{\ell'}||$ $\geq$ $u$.
Let $\mlor$ be an $(h,t)$-ambiguous function such that $\mlor(\oor
\cup (\bigcup_{\ell \in [r]} A_{\ell}))$ is defined for every
$A_{\ell} \subseteq U_{\ell}$ satisfying $||A_{\ell}|| \leq h$. If
$u > (3 \cdot d + d') \cdot r \cdot 2^{2r} \cdot p(m) \cdot
t(p(m))$, then there exist strings $\alpha_1 \in U_1$, $\alpha_2 \in
U_2$, $\ldots$, $\alpha_{r} \in U_{r}$ such that the following
properties hold:
\begin{description}
\item[(A)] For every $x \in X$ and for every pair of distinct, nonempty sets
$S_1$, $S_2 \in \powerset(\{1, 2, \ldots, r\})$, if the conditions:
\begin{description}
\item[(A.1)] $N^{\mlor(\oor \cup (\bigcup_{\ell \in S_1} \{\alpha_{\ell}\}))}(x)$
accepts and for all $S'_1 \subset S_1$, $N^{\mlor(\oor \cup
(\bigcup_{\ell \in S'_1} \{\alpha_{\ell}\}))}(x)$ rejects, and
\item[(A.2)] $N^{\mlor(\oor \cup (\bigcup_{\ell \in S_2} \{\alpha_{\ell}\}))}(x)$
accepts and for all $S'_2 \subset S_2$, $N^{\mlor(\oor \cup
(\bigcup_{\ell \in S'_2} \{\alpha_{\ell}\}))}(x)$ rejects,
\end{description}
are satisfied, then there are at least two accepting paths in
$N^{\mlor(\oor \cup (\bigcup_{\ell \in S_1 \cup S_2}
\{\alpha_{\ell}\}))}(x)$.
\item[(B)] For every $y \in Y$ and for every nonempty set
$S \in \powerset(\{1, 2, \ldots, r\})$, the following is true:
\[
\mbox{ if } N^{\mlor(\oor)}(y) \mbox{ accepts, then } N^{\mlor(\oor
\cup (\bigcup_{\ell \in S} \{\alpha_{\ell}\}))}(y) \mbox{ also
accepts.}
\]
\end{description}
\end{lemma}
\sproof We prove the lemma using the probabilistic method. Choose
$\alpha_1 \in U_1$, $\alpha_2 \in U_2$, $\ldots$, $\alpha_{r} \in
U_{r}$ uniformly and independently at random. Let $E_1$ be the event
that there exist an $x \in X$ and distinct, nonempty sets $S_1, S_2
\in \powerset(\{1, 2, \ldots, r\})$ satisfying the conditions (A.1)
and (A.2) given in the lemma, but $N^{\mlor(\oor \cup (\bigcup_{\ell
\in S_1 \cup S_2} \{\alpha_{\ell}\}))}(x)$ has at most one accepting
path. Let $E_2$ be the event that there exist a $y \in Y$ and a
nonempty set $S \in \powerset(\{1, 2, \ldots, r\})$ such that
$N^{\mlor(\oor)}(y)$ accepts, but $N^{\mlor(\oor \cup (\bigcup_{\ell
\in S} \{\alpha_{\ell}\}))}(y)$ rejects. We will prove that
$\prob(E_1) + \prob(E_2) < 1$, thus completing the proof of the
lemma.

We first make the following claim.

\begin{claim} \label{claim:s-not-changed}
Fix a string $z \in \sigmastar$. Let $\oor$, $U_1$, $U_2$, $\ldots$,
$U_r$ be sets defined in the statement of
Lemma~\ref{lemma:existential-strings}. Let $\mlor$ be an
$(h,t)$-ambiguous function such that $\mlor(\oor \cup (\bigcup_{\ell
\in [r]} A_{\ell}))$ is defined for every $A_{\ell} \subseteq
U_{\ell}$ satisfying $||A_{\ell}|| \leq h$. Let $T \subseteq [r]$.
Fix $\alpha_i \in U_i$ for each $i \in T$ arbitrarily. If we choose
$\alpha_j \in U_j$, for each $j \in [r] - T$, uniformly and
independently at random, then for any $T_1$, $T_2 \subseteq [r]$
satisfying $(T_1 \Delta T_2) \cap T = \emptyset$,
\[
z \in \mlor(\mathcal{O} \cup (\bigcup_{\ell \in T_1}
\{\alpha_{\ell}\})) \Longleftrightarrow z \notin \mlor(\mathcal{O}
\cup (\bigcup_{\ell \in T_2} \{\alpha_{\ell}\}))
\]
is true with probability $\le r\cdot t(|z|)/u$.
\end{claim}

\noindent {\bf Proof of Claim~\ref{claim:s-not-changed}.} \quad Let
$V = T_1 \cap T_2$. Because $\mlor$ is $(h,t)$-ambiguous, for any $j
\in T_1 \Delta T_2$, the probability over uniform random choice of
$\alpha_j \in U_j$ that
\[
z \in \mlor(\mathcal{O} \cup (\bigcup_{\ell \in V}
\{\alpha_{\ell}\})) \Longleftrightarrow z \notin \mlor(\mathcal{O}
\cup (\bigcup_{\ell \in V} \{\alpha_{\ell}\}) \cup \{\alpha_j\})
\]
is at most $t(|z|)/u$. Successively choose $\alpha_j \in U_j$
uniformly at random, where $j \in T_1 \Delta T_2$, and add $j$ to
$V$ until $V$ equals $T_1$. The probability that
\[
z \in \mlor(\mathcal{O} \cup (\bigcup_{\ell \in T_1}
\{\alpha_{\ell}\})) \Longleftrightarrow z \notin \mlor(\mathcal{O}
\cup (\bigcup_{\ell \in T_1 \cap T_2} \{\alpha_{\ell}\}))
\]
is, therefore, at most $||T_1 - T_2|| \cdot t(|z|)/u$. Likewise, the
probability that
\[
z \in \mlor(\mathcal{O} \cup (\bigcup_{\ell \in T_2}
\{\alpha_{\ell}\})) \Longleftrightarrow z \notin \mlor(\mathcal{O}
\cup (\bigcup_{\ell \in T_1 \cap T_2} \{\alpha_{\ell}\}))
\]
is at most $||T_2 - T_1|| \cdot t(|z|)/u$. Hence the probability
that
\[
z \in \mlor(\mathcal{O} \cup (\bigcup_{\ell \in T_1}
\{\alpha_{\ell}\})) \Longleftrightarrow z \notin \mlor(\mathcal{O}
\cup (\bigcup_{\ell \in T_2} \{\alpha_{\ell}\}))
\]
is at most $(||T_1 - T_2|| + ||T_2 - T_1||) \cdot t(|z|)/u$ $\leq$
$r \cdot t(|z|)/u$.~\qed(Claim~\ref{claim:s-not-changed})

\bigskip

\noindent Using Claim~\ref{claim:s-not-changed}, we obtain an upper
bound on $\prob(E_1)$ in Claim~\ref{claim:probE1-bound}, and an
upper bound on $\prob(E_2)$ in Claim~\ref{claim:probE2-bound}.

\begin{claim} \label{claim:probE1-bound}
$\prob(E_1) \leq 3 \cdot d \cdot r \cdot 2^{2r} \cdot p(m) \cdot
t(p(m))/u$.
\end{claim}

\noindent {\bf Proof of Claim~\ref{claim:probE1-bound}.} \quad Let
$C_{x, S_1, S_2}$ stand for the condition ``input $x$ and the pair
$S_1$, $S_2$ satisfy the conditions (A.1) and (A.2) given in the
lemma.'' Fix an $x \in X$ and distinct, nonempty sets $S_1$ and
$S_2$. Let $E_{x, S_1, S_2}$ denote the event that $C_{x, S_1, S_2}$
is satisfied, but $N^{\mlor(\oor \cup (\bigcup_{\ell \in S_1 \cup
S_2} \{\alpha_{\ell}\}))}(x)$ has at most one accepting path.
Clearly, $\prob(E_1) \leq \sum_{x, S_1, S_2}
\prob(E_{x, S_1, S_2})$. If condition $C_{x, S_1, S_2}$ is satisfied, then %
for each $i \in \{1,2\}$, we can fix an accepting path $\rho(S_i)$
in $N^{\mlor(\mathcal{O} \cup (\bigcup_{\ell \in
S_i}\{\alpha_{\ell}\}))}(x)$. For definiteness, let $\rho(S_i)$ be
the lexicographically first accepting path in $N^{\mlor(\mathcal{O}
\cup (\bigcup_{\ell \in S_i}\{\alpha_{\ell}\}))}(x)$.

If $E_{x, S_1, S_2}$ occurs, then at least one of the events $J$,
$H(S_1)$, or $H(S_2)$ occurs, where
\begin{itemize}
\item
$J$ is the event that condition $C_{x, S_1, S_2}$ is satisfied, and
$\rho(S_1)$ and $\rho(S_2)$ are equal.
\item
$H(S_1)$ is the event that condition $C_{x, S_1, S_2}$ is satisfied
and $\rho(S_1)$ does not appear in $N^{\mlor(\mathcal{O} \cup
(\bigcup_{\ell \in S_1\cup S_2}\{\alpha_{\ell}\}))}(x)$.
\item
$H(S_2)$ is the event that condition $C_{x, S_1, S_2}$ is satisfied
and $\rho(S_2)$ does not appear in $N^{\mlor(\mathcal{O} \cup
(\bigcup_{\ell \in S_1\cup S_2}\{\alpha_{\ell}\}))}(x)$.
\end{itemize}

\noindent We first determine an upper bound on $\prob(J)$. To this
end, we determine  an upper bound on $\prob(J \;|\; \alpha_i =
\beta_i \mbox{ for all $i\in S_1$})$ for arbitrary fixed strings
$\beta_i\in U_i$ for each $i\in S_1$. Hence, suppose henceforth that
$\alpha_i = \beta_i$ for all $i\in S_1$. Because $S_1 \neq S_2$ and
because we can assume that $S_1$ satisfies the condition (A.1) given
in the lemma, the (accepting) path $\rho(S_1)$ in
$N^{\mlor(\mathcal{O} \cup (\bigcup_{\ell \in
S_1}\{\alpha_{\ell}\}))}(x)$ does not appear in
$N^{\mlor(\mathcal{O} \cup (\bigcup_{\ell \in S_1\cap
S_2}\{\alpha_{\ell}\}))}(x)$. Hence for at least one string $z \in
Q(\rho(S_1))$ queried along $\rho(S_1)$, we have
\begin{equation} \label{eq:Gleichung1}
z \in \mlor(\mathcal{O} \cup (\bigcup_{\ell \in S_1\cap
S_2}\{\alpha_{\ell}\}))\; \Longleftrightarrow z \notin
\mlor(\mathcal{O} \cup (\bigcup_{\ell \in S_1}\{\alpha_{\ell}\})).
\end{equation}
Note that this condition depends only on strings $\alpha_i$ with
$i\in S_1$, which are fixed by $\alpha_i = \beta_i$. Fix one string
$z$ satisfying Statement~(\ref{eq:Gleichung1}). Applying
Claim~\ref{claim:s-not-changed} with $T := S_1$, $T_1 := S_1 \cap
S_2$, and $T_2 := S_2$, we get that
\begin{equation}\label{eq:Gleichung2}
z \in \mlor(\mathcal{O} \cup (\bigcup_{\ell \in S_1\cap
S_2}\{\alpha_{\ell}\}))\; \Longleftrightarrow z \in
\mlor(\mathcal{O} \cup (\bigcup_{\ell \in S_2}\{\alpha_{\ell}\}))
\end{equation}
holds with probability $\ge 1 - r\cdot t(|z|)/u \ge 1 - r\cdot
t(p(m))/u$. Statements~(\ref{eq:Gleichung1})
and~(\ref{eq:Gleichung2}) together imply that with probability $\ge
1 - r \cdot t(p(m))/u$,
\begin{equation}
z \in \mlor(\mathcal{O} \cup (\bigcup_{\ell \in
S_1}\{\alpha_{\ell}\})) \Longleftrightarrow z \notin
\mlor(\mathcal{O} \cup (\bigcup_{\ell \in S_2}\{\alpha_{\ell}\})).
\end{equation}
Hence with probability $\ge 1 - r \cdot t(p(m))/u$, path $\rho(S_1)$
does not appear in $N^{\mlor(\mathcal{O} \cup (\bigcup_{\ell \in
S_2}\{\alpha_{\ell}\}))}(x)$, and therefore $\rho(S_1) \neq
\rho(S_2)$. We have proven that for arbitrary fixed strings
\{$\beta_i\in U_i \condition i\in S_1$\}, it holds that $\prob(J
\;|\; \alpha_i = \beta_i \mbox{ for all $i\in S_1$}) \le r \cdot
t(p(m))/u$. Therefore, also $\prob(J) \le r \cdot t(p(m))/u$.

To determine an upper bound on $\prob(H(S_1))$, we determine  an
upper bound on $\prob(H(S_1) \;|\; \alpha_i = \beta_i \mbox{ for all
$i\in S_1$})$ for arbitrary fixed strings $\beta_i\in U_i$ for each
$i\in S_1$. Hence, suppose henceforth that $\alpha_i = \beta_i$ for
all $i\in S_1$. Clearly, $\rho(S_1)$ depends only on strings
$\alpha_i$ with $i\in S_1$, which we have fixed by $\alpha_i =
\beta_i$. Then the event $H(S_1)$ occurs only if it holds that
$\rho(S_1)$ queries some string $z$ with
\[
z \in \mlor(\mathcal{O} \cup (\bigcup_{\ell \in
S_1}\{\alpha_{\ell}\}))\; \Longleftrightarrow z \notin
\mlor(\mathcal{O} \cup (\bigcup_{\ell \in S_1 \cup
S_2}\{\alpha_{\ell}\})).
\]
Note that path $\rho(S_1)$ in $N^{\mlor(\mathcal{O} \cup
(\bigcup_{\ell \in S_1}\{\alpha_{\ell}\}))}(x)$ queries at most
$p(m)$ strings, each of which is of length $\leq p(m)$. Applying
Claim~\ref{claim:s-not-changed} with $T := S_1$, $T_1 := S_1$, and
$T_2 := S_1 \cup S_2$, we get that $\prob(H(S_1) \;|\; \alpha_i =
\beta_i \mbox{ for all $i\in S_1$}) \leq p(m) \cdot r \cdot
t(p(m))/u$. We have thus proven that for arbitrary fixed strings
$\{\beta_i\in U_i\condition i\in S_1\}$, it
  holds that
$\prob(H(S_1) \;|\; \alpha_i = \beta_i \mbox{ for all $i\in S_1$})
\le p(m) \cdot r \cdot t(p(m))/u$. Therefore, also $\prob(H(S_1))
\le p(m) \cdot r \cdot t(p(m))/u$.

Analogously, we can prove that $\prob(H(S_2)) \leq p(m) \cdot r
\cdot t(p(m))/u$.

Thus $\prob(E_{x, S_1, S_2}) \leq \prob(J \cup H(S_1) \cup H(S_2))
\leq \prob(J) + \prob(H(S_1)) + \prob(H(S_2)) \leq 3 \cdot p(m)
\cdot r \cdot t(p(m))/u$. It follows that $\prob(E_1) \leq 3 \cdot d
\cdot r \cdot 2^{2r} \cdot p(m) \cdot
t(p(m))/u$.~\qed(Claim~\ref{claim:probE1-bound})

\bigskip

\begin{claim} \label{claim:probE2-bound}
$\prob(E_2) \leq d' \cdot r \cdot 2^{r} \cdot p(m) \cdot t(p(m))/u$.
\end{claim}

\noindent {\bf Proof of Claim~\ref{claim:probE2-bound}.} Fix a $y
\in Y$ and a nonempty set $S \in \powerset(\{1, 2, \ldots, r\})$.
Let $E_{y, S}$ denote the event that $N^{\mlor(\oor)}(y)$ accepts,
but $N^{\mlor(\oor \cup (\bigcup_{\ell \in S}
\{\alpha_{\ell}\}))}(y)$ rejects. Clearly, $\prob(E_2) \leq \sum_{y,
S} \prob(E_{y,S})$. If $N^{\mlor(\oor)}(y)$ accepts, then we can fix
an accepting path $\rho$ in $N^{\mlor(\oor)}(y)$. For definiteness,
let $\rho$ be the lexicographically first accepting path in
$N^{\mlor(\oor)}(y)$.

Suppose $E_{y,S}$ occurs. Since $N^{\mlor(\oor \cup (\bigcup_{\ell
\in S} \{\alpha_{\ell}\}))}(y)$ rejects, the (accepting) path $\rho$
in $N^{\mlor(\oor)}(y)$ does not appear in $N^{\mlor(\oor \cup
(\bigcup_{\ell \in S} \{\alpha_{\ell}\}))}(y)$. Hence for at least
one string $z \in Q(\rho)$ queried along $\rho$, we have
\[
z \in \mlor(\mathcal{O})\; \Longleftrightarrow z \not \in \mlor(\oor
\cup (\bigcup_{\ell \in S}\{\alpha_{\ell}\})).
\]
Applying Claim~\ref{claim:s-not-changed} with $T := \emptyset$, $T_1
:= \emptyset$, and $T_2 := S$, we get that for each $z \in Q(\rho)$,
\[
z \in \mlor(\mathcal{O})\; \Longleftrightarrow z \not \in \mlor(\oor
\cup (\bigcup_{\ell \in S}\{\alpha_{\ell}\}))
\]
holds with probability $\leq r \cdot t(|z|)/u$. Since $||Q(\rho)||
\leq p(m)$ and since the length of each query $z \in Q(\rho)$ is at
most $p(m)$, with probability $\leq p(m) \cdot r \cdot t(p(m))/u$ we
have, for some $z \in Q(\rho)$,
\[
z \in \mlor(\mathcal{O})\; \Longleftrightarrow z \not \in \mlor(\oor
\cup (\bigcup_{\ell \in S}\{\alpha_{\ell}\})).
\]
Thus we have shown that $\prob(E_{y,S}) \leq p(m) \cdot r \cdot
t(p(m))/u$. It follows that $\prob(E_2) \leq d' \cdot r \cdot 2^{r}
\cdot p(m) \cdot t(p(m))/u$.~\qed(Claim~\ref{claim:probE2-bound})

\bigskip

\noindent From Claim~\ref{claim:probE1-bound} and
Claim~\ref{claim:probE2-bound}, we get $\prob(E_1)$ $+$ $\prob(E_2)$
$\leq$ $(3 \cdot d + d') \cdot r \cdot 2^{2r} \cdot p(m) \cdot
t(p(m))/u < 1$, by our choice of $u$. This completes the
proof.~\qed(Lemma~\ref{lemma:existential-strings}) \\

\noindent Now it is relatively easy to prove the main result of this subsection.

\bigskip

\noindent {\bf Proof of Theorem~\ref{thm:promis-truth-table-versus-Turing}.\quad}
For each length $n$, we will reserve the following segment of
$k(n)+1$ regions $S_{n,f} = 1^n01^f0\Sigma^{3k(n)+n}$, where $f\in
[k(n)+1]$. For $n\ge 1$, define $S_n = \bigcup_{f=1}^{k(n)+1}
S_{n,f}$. We take the test language $L(\aor)$ used in the proof of
Theorem~\ref{thm:truth-table-versus-Turing}. %
The oracle $\aor$ is constructed in stages. Let
$(N_{i,1},N_{i,2},\ldots , N_{i,h} , M_j)$ be an enumeration of
tuples where $N_{i, *}$ is a nondeterministic polynomial-time oracle
Turing machine, and $M_j$ is a deterministic polynomial-time oracle
Turing machine making, for any set $\aor$ and any input of length
$n$, at most $k(n)$ queries to $L[\aor; N_{i,1}, N_{i,2}, \ldots ,
N_{i,h}]$ and at most polynomially many queries to $\aor$. Initially
$\aor := \emptyset$.

\medskip

\noindent {\bf Stage \boldmath $\langle i, j\rangle$:\unboldmath }
Let $p(\cdot)$ be a polynomial that bounds the running time of both
$N_{i, \star}$ and $M_j$. Choose a very large integer $n$ such that
(a) $2^{3k(n)+n} - p(n) > 4 \cdot k(n)\cdot (k(n)+1)\cdot
2^{2(k(n)+1)}\cdot p(p(n)) \cdot t(p(p(n)))$, where $t(\cdot)$ is a
polynomial defined later in this proof, (b) no string of length $n$
or more is queried in any of the previous stages, (c) $n$ is larger
than the value in the previous stage, and (d) $n$ satisfies any
promises made in the previous stages.

If there exists a set $B\subseteq S_n$, satisfying $||B\cap
S_{n,f}|| \le h$ for every $f\in [k(n)+1]$, such that (a) $[\aor
\cup B; N_{i,2}, N_{i,3}, \ldots, N_{i,h}]$ is not unambiguous, or
(b) $M_{j}(0^n)$ queries $\beta$ to $[\aor \cup B; N_{i,1}, N_{i,2},
\ldots, N_{i,h}]$ and $N_{i,1}^{L[\aor \cup B; N_{i,2}, N_{i,3},
\ldots, N_{i,h}]}(\beta)$ is not unambiguous, then set $\aor := \aor
\cup B$. Promise to choose the value of $n$ in the next stage to be
sufficiently large so that any of the requirements $(a)$, $(b)$
satisfied in this stage cannot become invalid in the next stage.

Otherwise if no such set $B \subseteq S_n$ exists, then we proceed
as follows. Define an $(h,t)$-ambiguous function $\mlor:
\powerset(\sigmastar) \rightarrow \powerset(\sigmastar)$ so that for
every $\oor \subseteq\sigmastar$,
\[
  \mlor(\oor) =_{df} L_{\textnormal{unambiguous}}[\oor ; N_{i,2}, N_{i,3}, \ldots ,
  N_{i,h}],
\]
where $t(\cdot)$ is some polynomial.\footnote{As observed in the
proof of Theorem~\ref{thm:truth-table-versus-Turing}, we can take
$t(\cdot)$ to be the polynomial $5^{h-1} \cdot \prod_{i=1}^{h-1} (p
\circ)^{i}(\cdot)$.} It is easy to see that $\mlor(\aor \cup B)$ is
defined for every $B \subseteq S_n$, which satisfies $||B \cap
S_{n,f}|| \leq h$ for every $f \in [k(n) + 1]$. We next use
Claim~\ref{claim:idenitical-comp} to successfully finish this stage.
\begin{claim}
\label{claim:idenitical-comp} There is a string $\alpha \in S_n$
such that $M_{j}(0^n)$ with oracle $\aor \oplus
L(N_{i,1}^{\mlor(\aor)})$ is identical to $M_{j}(0^n)$ with oracle
$(\aor \cup \{\alpha\}) \oplus L(N_{i,1}^{\mlor(\aor \cup
\{\alpha\})})$.
\end{claim}
That is, if $M_{j}(0^n)$ with oracle $\aor \oplus
L(N_{i,1}^{\mlor(\aor)})$ rejects, then we set $\aor := \aor \cup
\{\alpha\}$; otherwise if $M_{j}(0^n)$ with oracle $\aor \oplus
L(N_{i,1}^{\mlor(\aor)})$ accepts, then we leave $\aor$ unchanged.
Finally, we move to the next stage. \\
{\bf End of Stage} \\

\noindent
This completes the proof of
Theorem~\ref{thm:promis-truth-table-versus-Turing}.~\qed~(Theorem~\ref{thm:promis-truth-table-versus-Turing})

\medskip

\noindent {\bf Proof of Claim~\ref{claim:idenitical-comp}.} \quad
Let $\beta_1$, $\beta_2$, $\ldots$, $\beta_{k(n)}$ be the sequence
of queries made by $M_{j}(0^n)$ to the oracle
$L(N_{i,1}^{\mlor(\aor)})$. Let $I = \{\ell \condition
N_{i,1}^{\mlor(\aor)}(\beta_{\ell}) \text{ accepts} \}$. Let
$\widetilde Q$ be the set of strings that are queried by
$M_{j}(0^n)$ to oracle $\aor$.
Clearly, $||\widetilde Q|| \le p(n)$.

Apply Lemma~\ref{lemma:existential-strings} with $N := N_{i,1}$,
$\oor := \aor$, $d := k(n)$, $d' := k(n)$, $X = \{\beta_1, \beta_2,
\ldots, \beta_{k(n)}\}$, $Y = \{\beta_1, \beta_2, \ldots,
\beta_{k(n)}\}$, $m:=p(n)$, $r: = k(n)+1$, $U_{f} := S_{n,f}-
\widetilde{Q}$ for each $f \in [k(n)+1]$, and $u := 2^{3k(n)+n} -
p(n)$. %
We obtain strings $\alpha_1 \in S_{n,1}-\widetilde{Q}$, $\alpha_2
\in S_{n,2}-\widetilde{Q}$, $\ldots$, $\alpha_{k(n)+1} \in
S_{n,k(n)+1}-\widetilde{Q}$, which %
satisfy the properties (A) and (B) given in
Lemma~\ref{lemma:existential-strings}. Now assign to each query
$\beta_{\ell}$ with $\ell\in [k(n)] - I$ a collection
$\mathcal{S}_{\ell} \subseteq \powerset([k(n)+1])$ in the following
way: $\{f_1, f_2, \ldots, f_s\} \in \mathcal{S}_{\ell}$ if and only
if
\begin{description}
\item[(a)] adding $\{\alpha_{f_1}, \alpha_{f_2}, \ldots, \alpha_{f_{s}}\}$
to $\aor$ makes $N_{i,1}^{\mlor(\aor)}(\beta_{\ell})$ change from
rejection to acceptance, i.e., $N_{i,1}^{\mlor(\aor)}(\beta_{\ell})$
rejects but $N_{i,1}^{\mlor(\aor \cup \{\alpha_{f_1}, \alpha_{f_2},
\ldots, \alpha_{f_{s}}\})}(\beta_{\ell})$ accepts, and
\item[(b)] no set $T \subset \{f_1, f_2, \ldots, f_{s}\}$
satisfies (a), i.e., for no such set
$T$ it holds that %
$N_{i,1}^{\mlor(\aor \cup (\bigcup_{j \in T }
\{\alpha_{j}\}))}(\beta_{\ell})$ accepts.
\end{description}
Note that no collection $\mathcal{S}_{\ell}$ contains the empty set.
However, some of these collections may be empty.

Suppose that Claim~\ref{claim:idenitical-comp} is not true. Then for
every $e\in [k(n)+1]$, there is an $\ell \in [k(n)]$ such that
\begin{equation} \label{eq:gl1}
N_{i,1}^{\mlor(\aor)}(\beta_\ell) \text{ rejects }
\Longleftrightarrow N_{i,1}^{\mlor(\aor \cup \{
\alpha_e\})}(\beta_\ell) \text{ accepts.}
\end{equation}
Because $\alpha_1$, $\alpha_2$, $\ldots$, $\alpha_{k(n) +1}$ satisfy
property (B) of Lemma~\ref{lemma:existential-strings}, Statement
~(\ref{eq:gl1}) can only be true for $\ell\notin I$. This implies
that for every $e\in [k(n)+1]$, there is an $\ell \in [k(n)] - I$
such that
\begin{equation*}
N_{i,1}^{\mlor(\aor)}(\beta_\ell) \text{ rejects } \text{ and }
N_{i,1}^{\mlor(\aor \cup \{ \alpha_e\})}(\beta_\ell) \text{
accepts.}
\end{equation*}
It follows from the definition of the collections
$\mathcal{S}_{\ell}$ that for every $e\in [k(n)+1]$, there is a
collection $\mathcal{S}_{\ell} $ such that the singleton $\{
\alpha_e \}$ is contained in $\mathcal{S}_{\ell}$. Thus we have
proven condition (1) of Lemma~\ref{lemma:gaming-lemma} (the gaming
lemma).

Now take two distinct sets $A,B\in \mathcal{S}_\ell$ for some
$\ell\in [k(n)]$. Then by the definition of the collections
$\mathcal{S}_\ell$ together with property (A) of
Lemma~\ref{lemma:existential-strings}, $N_{i,1}^{\mlor(\aor \cup
(\bigcup_{e \in A \cup B } \{\alpha_{e}\}))}(\beta_{\ell})$ has at
least two accepting paths. Because of our assumption about the
unambiguity, we can be sure that adding $\bigcup_{e \in A \cup B }
\{\alpha_{e}\}$ to $\aor$ changes the decision (i.e, acceptance or
rejection) of $N_{i,1}^{\mlor(\aor)}$  for a previous query
$\beta_{\ell'}$ with $\ell' < \ell$. The decision of
$N_{i,1}^{\mlor(\aor)}(\beta_{\ell'})$ %
on addition of $\bigcup_{e \in A \cup B } \{\alpha_{e}\}$ to $\aor$
must change from rejection to acceptance, and not from acceptance to
rejection, because $\alpha_1$, $\alpha_2$, $\ldots$, $\alpha_{k(n)
+1}$ satisfy property (B) of Lemma~\ref{lemma:existential-strings}.
Hence there is a set $C \in \mathcal{S}_{\ell'}$ such that $C
\subseteq A\cup B$. This proves condition (2) of
Lemma~\ref{lemma:gaming-lemma}. Lemma~\ref{lemma:gaming-lemma}
implies that the number of queries $k(n)$ is greater than or equal
to the number of regions $k(n)+1$, a
contradiction.~\qed~(Claim~\ref{claim:idenitical-comp})

\subsection{Simulating Adaptive Access by Nonadaptive Access}
\label{subsec:adap-nonadap}

\noindent Sections~\ref{subsec:nonadap-adap-nonpromise}
and~\ref{subsec:nonadap-adap-promise} studied the limitations of
simulating nonadaptive queries to $\up_{\leq h}$ by adaptive queries
to ${{\rm U}}\Sigma^{p}_{h}$ in relativized settings. This section
complements these investigations. In particular,
Corollary~\ref{cor:up-up-constant-kh} of this section shows that in
a certain relativized world, it is impossible to simulate adaptive
$k$-Turing access to $\up_{\leq h}$ by nonadaptive $(2^k-2)$-tt
access to ${{\rm U}}\Sigma^{p}_{h}$. This also implies optimality of
robustly (i.e., for every oracle) simulating adaptive $k$-Turing
accesses by nonadaptive $(2^k-1)$-tt accesses to classes such as
$\up_{\leq h}$ and ${{\rm U}}\Sigma^{p}_{h}$, since for any class
$\mathcal{C}$, we can easily, via a brute-force method, simulate
adaptive $k$-Turing reduction to $\mathcal{C}$ by nonadaptive
$(2^k-1)$-tt reduction to $\mathcal{C}$.

The proof of Theorem~\ref{thm:const-kh-Ttt} employs a technique of
Buhrman, Spaan, and Torenvliet~\cite{buh-spa-tor:b:bounded}, which
Cai, Hemachandra, and Vysko\v{c}~\cite{cai-hem-vys:b:promise}
referred to by ``force your way through the tree" technique.
Buhrman, Spaan, and Torenvliet~\cite{buh-spa-tor:b:bounded} used
their technique to prove that $\nexp$ has a set that is complete for
$k$-Turing reductions, but not complete for $(2^k-2)$-tt reductions.
Cai, Hemachandra, and Vysko\v{c}~\cite{cai-hem-vys:b:promise} used
this technique to prove Theorem~\ref{thm:const-kh-Ttt} for the case
of $h=1$. We use the same approach to generalize the result of Cai,
Hemachandra, and Vysko\v{c}~\cite{cai-hem-vys:b:promise} from the
case of $h=1$ to the case of arbitrary integer $h \geq 1$.

\begin{theorem}
\label{thm:const-kh-Ttt} For any integers $k, h \geq 1$, there
exists an oracle $\aor$ such that
\[
R^{p}_{k\hbox{-}T}(\up_{\leq h}^{\aor}) \nsubseteq
R^{p, \aor}_{(2^{k}-2)\hbox{-}tt}(\np^{{{\rm U}}\Sigma_{h-1}^{p,\aor}}).
\]
\end{theorem}
\sproof For each length $n$, we will reserve the following segment
of $2^k-1$ regions: $S_{n,f} = 1^n0f0\Sigma^n$, where $f \in
(\Sigma^*)^{\leq k-1}$. For $n \geq 1$, define $S_n = \bigcup_{f \in
(\Sigma^*)^{\leq k-1}} S_{n,f}$. For each length $n$ and set $A
\subseteq \sigmastar$, we also define a sequence
$b^{A}_{n,1}b^{A}_{n,2}\ldots b^{A}_{n,k}$ of bits as follows: \\ \\
$(\star)$
\[
b^{A}_{n,1} = \left\{
\begin{array}{ll}
1 & \mbox{if $S_{n,\epsilon} \bigcap A \neq \emptyset$,}  \\
0 & \mbox{otherwise,}
\end{array}
\right.
\]
and for each $\ell$ with $2 \leq \ell \leq k$,
\[
b^{A}_{n,\ell} = \left\{
\begin{array}{ll}
1 & \mbox{if $S_{n,b^{A}_{n,1}b^{A}_{n,2}\ldots b^{A}_{n,\ell-1}} \bigcap A \neq \emptyset$,} \\
0 & \mbox{otherwise.}
\end{array}
\right.
\]
Our test language is $L(\aor) = \{0^n~|~b^{\aor}_{n,k} = 1\}$. Note
that our test language is the same as the one
in~\cite[Theorem~3.3]{cai-hem-vys:b:promise}. We stipulate that for
all $n \geq 1$ and for all $f \in (\Sigma^*)^{\leq k-1}$,
$||S_{n,f}|| \leq h$. Clearly, if the oracle set $\aor$ maintains
this stipulation, then we have $L(\aor) \in
R^{p}_{k\hbox{-}T}(\up_{\leq h}^{\aor})$. We construct an oracle
$\aor$ such that $L(\aor) \not \in
R^{p}_{(2^{k}-2)\hbox{-}tt}(\np^{{{\rm U}}\Sigma_{h-1}^{p,\aor}})$.
The construction can be easily modified to prove the stronger result
that $L(\aor) \not \in R^{p, \aor}_{(2^{k}-2)\hbox{-}tt}(\np^{{{\rm
U}}\Sigma_{h-1}^{p,\aor}})$.

Let $(N_{i,1},N_{i,2},\ldots , N_{i,h} , M_j)$ be an enumeration of
tuples, where $N_{i, *}$ is a nondeterministic polynomial-time oracle
Turing machine, and $M_j$ is a deterministic polynomial-time oracle
Turing machine making, for any set $\aor$ and for any input, at most
$2^k-2$ nonadaptive queries to $L[\aor; N_{i,1}, N_{i,2}, \ldots ,
N_{i,h}]$. %
Initially, let $\aor := \emptyset$.

\medskip

\noindent {\bf Stage \boldmath $\langle i, j\rangle$:\unboldmath }
Let $p(\cdot)$ be a polynomial that bounds the running time of both
$N_{i, \star}$ and $M_j$. Choose a very large integer $n$ such that
(a) $2^{n}  > 2^{2k} \cdot p(p(n)) \cdot t(p(p(n)))$, where $t(\cdot)$ is a
polynomial defined later in this proof, (b) no string of length $n$
or more is queried in any of the previous stages, (c) $n$ is larger
than the value in the previous stage, and (d) $n$ satisfies any
promises made in the previous stages.

If there exists a set $B \subseteq S_n$ satisfying $||B \cap
S_{n,f}|| \leq h$ for every $f \in (\Sigma^*)^{\leq k-1}$ such that
$[\aor \cup B; N_{i,2}, \ldots, N_{i,h}]$ is not unambiguous, then
set $\aor := \aor \cup B$. Promise to choose the value of $n$ in the
next stage to be larger than $(p\circ)^{h-1}(|w|)$, where $w$ is an
arbitrary string witnessing that $[\aor \cup B; N_{i,2}, \ldots,
N_{i,h}]$ is not unambiguous, and then move to the next stage.

Otherwise, we define  a %
function $\mlor : \powerset(\sigmastar) \rightarrow
\powerset(\sigmastar)$ as follows: For every $\oor \subseteq
\sigmastar$, let
\[
\mlor(\oor) =_{df} L_{\textnormal{unambiguous}}[\oor ; N_{i,2},
N_{i,3}, \ldots , N_{i,h}].
\]
It is easy to see that by Lemma~\ref{lemma:weakness-uph-machines},
$\mlor$ is $(h,t)$-ambiguous for polynomial $t(.) =_{df}
5^{h-1}\cdot \prod_{\ell=1}^{h-1}(p\circ)^{\ell}(\cdot)$. Also,
$\mlor(\aor \cup B)$ is defined for every $B \subseteq S_n$
that satisfies $||B \cap S_{n,f}|| \leq h$ for every $f \in
(\Sigma^*)^{\leq k-1}$.

Let $\beta_1$, $\beta_2$, $\ldots$, $\beta_{2^k-2}$ be the sequence
of nonadaptive queries made by $M_j(0^n)$ to the oracle
$L(N_{i,1}^{\mlor(\aor)})$. Consider the following procedure
\emph{Diagonalize}.

\begin{center}
\begin{boxedminipage}[b]{6in}\small
\begin{tabbing}
\TABSET {\bf \boldmath Procedure Diagonalize($\{\alpha_{f} \in S_{n,f}~|~f \in (\Sigma^*)^{\leq k-1}\}$)} \\
1. \> $\aor_{1}$ := $\aor$; \\ %
2. \> For $t := 1$ to $2^k-1$ do \\
3. \> \> Let $b_{n,1}^{\aor_t}b_{n,2}^{\aor_t}\ldots
b_{n,k}^{\aor_t}$ be the bit sequence given by $(\star)$;\\
4. \> \> If ($b_{n,k}^{\aor_t} = 0 \Longleftrightarrow
M_{j}^{L(N_{i,1}^{\mlor(\aor_t)})}(0^n)$ accepts) is true then \\
5. \> \> \> Output $\aor_t$ and terminate; \\
6. \> \> Else /* That is, $b_{n,k}^{\aor_t} = 0 \Longleftrightarrow
M_{j}^{L(N_{i,1}^{\mlor(\aor_t)})}(0^n)$ rejects */ \\
\> \> \> (6.1) \> \> Let $s := \max\{\ell \in [k]~|~b^{\aor_t}_{n,\ell} = 0\}$; \\
\> \> \> (6.2) \> \> $\aor_{t+1} := \aor_{t} \cup
\{\alpha_{b_{n,1}^{\aor_t}b_{n,2}^{\aor_t}\ldots
b_{n,s-1}^{\aor_t}}\}$; /* That is, flip $b_{n,k}^{\aor_t}$. */ \\
\> \> \> (6.3) \> \> If for every query
$\beta_{\ell}$, it holds that if $N_{i,1}^{\mlor(\aor_{t})}(\beta_{\ell})$ rejects
then \\
\> \> \> \> \> $N_{i,1}^{\mlor(\aor_{t+1})}(\beta_{\ell})$ also rejects, then \\
\> \> \> \> \> \> Output $\aor_{t+1}$ and terminate. \\
\> \> \> (6.4) \> \> Else \> \> /* \> there is a query $\beta_{\ell}$ such
that $N_{i,1}^{\mlor(\aor_t)}(\beta_{\ell})$ \\
\> \> \> \> \> \> \> \> rejects, but $N_{i,1}^{\mlor(\aor_{t+1})}(\beta_{\ell})$ accepts. */ \\
\> \> \> \> \> \> Return to the for loop;
\end{tabbing}
\textbf{End of Procedure}
\end{boxedminipage}
\end{center}

\medskip

\begin{claim}
\label{claim:exist-alpha-forf} For each $f \in (\Sigma^*)^{\leq
k-1}$, there exists $\widehat{\alpha_{f}} \in S_{n,f}$ such that for
each $t \in [2^k-1]$ and for each $\ell \in [2^k-2]$, the following
holds in the execution of Diagonalize($\{\widehat{\alpha_{f}}~|~f
\in (\Sigma^*)^{\leq k-1}\}$):
\[
\mbox{if $N_{i,1}^{\mlor(\aor_t)}(\beta_{\ell})$ accepts, then
$N_{i,1}^{\mlor(\aor_{t+1})}(\beta_{\ell})$ also accepts.}
\]
\end{claim}
Let us assume that Claim~\ref{claim:exist-alpha-forf} is true. Then
there exist strings $\widehat{\alpha_f} \in S_{n,f}$, for each $f
\in (\Sigma^*)^{\leq k-1}$, satisfying the property stated in the
claim. Set $\aor : =$ Diagonalize($\{\widehat{\alpha_{f}}~|~f \in
(\Sigma^*)^{\leq k-1}\}$)
and move to the next stage. \\
{\bf End of Stage}

\medskip

\noindent
For each $f \in (\Sigma^*)^{\leq k-1}$, let $\widehat{\alpha_{f}}
\in S_{n,f}$ be the strings promised in that claim. Notice that the
procedure Diagonalize($\{\widehat{\alpha_{f}}~|~f \in
(\Sigma^*)^{\leq k-1}\}$) never adds more than one string in any
region $S_{n,f}$. This follows because each region $S_{n,f}$ is
associated with exactly one string $\widehat{\alpha_{f}} \in
S_{n,f}$, and only these associated strings are ever considered for
inclusion in the oracle. Also note that the effect of Step (6.2) in
the procedure is to increment the binary number
$b_{n,1}^{\aor_t}b_{n,2}^{\aor_t}\ldots b_{n,k}^{\aor_t}$ by $1$.
That is, we have for each $t \in [2^k-1]$ considered until the
termination of the for loop,
$b_{n,1}^{\aor_{t+1}}b_{n,2}^{\aor_{t+1}}\ldots
b_{n,k}^{\aor_{t+1}}$ := $b_{n,1}^{\aor_t}b_{n,2}^{\aor_t}\ldots
b_{n,k}^{\aor_t} + 1$. This implies that after the execution of Step
(6.2), the bit $b_{n,k}^{\aor_t}$ is flipped, i.e.,
$b_{n,k}^{\aor_{t+1}} = \overline{b_{n,k}^{\aor_t}}$.

If Diagonalize($\{\widehat{\alpha_{f}}~|~f \in (\Sigma^*)^{\leq
k-1}\}$) terminates at Step 5, then clearly $0^n \in L(\aor)
\Longleftrightarrow$ $0^n \not \in
L(M_{j}^{L(N_{i,1}^{\mlor(\aor)})})$ and so we successfully finish
the stage. Otherwise, Diagonalize($\{\widehat{\alpha_{f}}~|~f \in
(\Sigma^*)^{\leq k-1}\}$) terminates at the execution of Step 6.3 or
it terminates because the for loop had finished iterating over the
range of values of $t$.

If Diagonalize($\{\widehat{\alpha_{f}}~|~f \in (\Sigma^*)^{\leq
k-1}\}$) terminates at the execution of Step 6.3, then we have the
following situation:

\begin{itemize}
\item $b_{n,k}^{\aor_t} = 0 \Longleftrightarrow
M_{j}^{L(N_{i,1}^{\mlor(\aor_t)})}(0^n)$ rejects.
\item $b_{n,k}^{\aor_{t+1}} = \overline{b_{n,k}^{\aor_t}}$.
\item For every query $\beta_{\ell}$, where $\ell \in [2^k-2]$, if
$N_{i,1}^{\mlor(\aor_t)}(\beta_{\ell})$ rejects, then $N_{i,1}^{\mlor(\aor_{t+1})}(\beta_{\ell})$ also
rejects.
\item For every query $\beta_{\ell}$, where $\ell \in [2^k-2]$, if
$N_{i,1}^{\mlor(\aor_t)}(\beta_{\ell})$ accepts then $N_{i,1}^{\mlor(\aor_{t+1})}(\beta_{\ell})$ also
accepts, by Claim~\ref{claim:exist-alpha-forf}.
\end{itemize}
It follows that $b_{n,k}^{\aor_{t+1}} = 1$ if and only if
$M_{j}^{L(N_{i,1}^{\mlor(\aor_{t+1})})}(0^n)$ rejects. Hence, we
successfully finish the stage. We next claim that if
Diagonalize($\{\widehat{\alpha_{f}}~|~f \in (\Sigma^*)^{\leq
k-1}\}$) does not terminate at Step 5, then it must terminate at the
execution of Step 6.3.

To this end, for each $t \in [2^k]$, let us define $Q_{\acc}(t)$ to
be the set of queries $\beta_{\ell}$ on which $N_{i,1}$ with oracle
$\mlor(\aor_t)$ accepts. Formally, for any $t \in [2^k]$, let
$Q_{\acc}(t) =_{df} \{\beta_{\ell}~|~\ell \in [2^k-2] \textnormal{
and } N_{i,1}^{\mlor(\aor_t)}(\beta_{\ell}) \textnormal{
accepts}\}$. By Claim~\ref{claim:exist-alpha-forf}, once a query
$\beta_{\ell}$ becomes a member of $Q_t$, the query $\beta_{\ell}$
remains accepted by $N_{i,1}^{\mlor(\aor_{t'})}$ for any $t \leq t'
\in [2^k]$. That is, $Q_{\acc}(t) \subseteq Q_{\acc}(t+1)$ for all
$t \in [2^k-1]$. By the definition of the sets $Q_{\acc}(t)$, it
follows that if the condition in Step (6.4) is true at some
iteration $t$ of the for loop, then there exist queries
$\beta_{\ell} \in Q_{\acc}(t+1) - Q_{\acc}(t)$; i.e., we have
$||Q_{\acc}(t+1)|| > ||Q_{\acc}(t)||$ at these iterations $t$. Thus,
there will be an iteration at which the condition in Step (6.4) will
not be true. (This follows because the number of iterations,
$(2^k-1)$, of the for loop is greater than the maximum possible
size, $2^k-2$, of $Q_{\acc}(t)$.) Therefore, at that iteration, the
condition in Step (6.3) will be true. Hence,
Diagonalize($\{\widehat{\alpha_{f}}~|~f \in (\Sigma^*)^{\leq
k-1}\}$) will terminate at the execution of Step 6.3. This completes
the proof of the theorem.~\qed~(Theorem~\ref{thm:const-kh-Ttt}) \\

\noindent {\bf Proof of Claim~\ref{claim:exist-alpha-forf}.} \quad
Let $f$ be arbitrary in $(\Sigma^*)^{\leq k-1}$. We will prove that
there is a small set $\widetilde{Q}(f)$ such that for each $t \in
[2^k-1]$ and for each $\ell \in [2^k-2]$, the following holds for
all $\alpha_f \in S_{n,f} - \widetilde{Q}(f)$:
\begin{equation}
\label{eq:scheisse} \mbox{if $N_{i,1}^{\mlor(\aor_t)}(\beta_{\ell})$
accepts, then $N_{i,1}^{\mlor(\aor_{t}\cup \alpha_f
)}(\beta_{\ell})$ also accepts.}
\end{equation}
To this end, fix $t \in [2^k-1]$ and $\ell \in [2^k-2]$ and assume
that $N_{i,1}^{\mlor(\aor_t)}(\beta_{\ell})$ accepts. Let $\rho$ be
an arbitrary accepting path $\rho$ in
$N_{i,1}^{\mlor(\aor_t)}(\beta_{\ell})$. Because $\mlor$ is
$(h,t)$-ambiguous, for each $z\in\sigmastar$ there can be at most
$t(|z|)$ strings $\alpha_f \in S_{n,f}$ such that
\begin{eqnarray*}
z \in \mlor(\aor_t) &\Longleftrightarrow& z \notin
\mlor(\aor_{t}\cup \alpha_f   ).
\end{eqnarray*}
Since there are at most $p(p(n))$ strings queried on $\rho$, there
is a set $\widetilde{Q}(f, t, \ell)$ with $||\widetilde{Q}(f, t,
\ell)|| \le p(p(n)) \cdot t(p(p(n)))$ ensuring that
$N_{i,1}^{\mlor(\aor_{t} \cup \alpha_f)}(\beta_{\ell})$ accepts for
every $\alpha_f \in S_{n,f} - \widetilde{Q}(f, t, \ell)$.

It is easy to see that Statement~(\ref{eq:scheisse}) is satisfied
with $\widetilde{Q}(f) = \bigcup_{t,\ell} \widetilde{Q}(f,t,\ell)$.
Clearly, $||\widetilde{Q}(f)|| \le (2^k-1)\cdot (2^k-2)\cdot p(p(n))
\cdot t(p(p(n)))< 2^n$ by our choice of $n$. Since
$||S_{n,f}||=2^n$, there exists a string $\widehat{\alpha_{f}} \in
S_{n,f}$ witnessing the correctness of
Claim~\ref{claim:exist-alpha-forf}.~\qed~(Claim~\ref{claim:exist-alpha-forf})

\begin{corollary}
\label{cor:up-up-constant-kh} For any integers $k, h \geq 1$, there
exists an oracle $\aor$ such that
\[
R^{p}_{k\hbox{-}T}(\up_{\leq h}^{\aor}) \nsubseteq
R^{p, \aor}_{(2^{k}-2)\hbox{-}tt}({{\rm U}}\Sigma_{h}^{p,\aor}).
\]
\end{corollary}

\subsection{Fault-tolerant Access}
\label{subsec:fault-tolerance}

\noindent Ko~\cite{ko:j:helping} introduced the notion of
\emph{one-sided helping} by a set $A$ in the computation of a set
$B$. A set $A$ is said to provide \emph{one-sided help} to a set $B$
if there is a deterministic oracle Turing machine $M$ computing $B$
and a polynomial $p(\cdot)$ such that (a) on any input $x \in B$,
$M^A(x)$ accepts in time $p(|x|)$, and (b) for all inputs $y$ and
for all oracles $C$, $M^{C}(y)$ accepts (though perhaps $M^C(y)$ may
take a longer time than $p(|y|)$) if and only if $y \in B$. Since
the machine $M$, accepting the set $B$, is capable of answering
correctly on faulty oracles, i.e., oracles $C$ different from the
oracle $A$ that provides one-sided help to $B$, the oracle access
mechanism is termed fault-tolerant
(see~\cite{cai-hem-vys:b:promise}). Ko~\cite{ko:j:helping} defined
$\p_{1\hbox{-}\help}(A)$ to be the class of all sets $B$ that can be
one-sided helped by $A$.

It is known that sets that can be one-sided helped (by any arbitrary
helper) are precisely those in $\np$~\cite{ko:j:helping}. Therefore,
the notion of one-sided helping provides an avenue for understanding
the structure of $\np$. For instance, given any class $\mathcal{C}
\subseteq \np$, what class of sets in $\np$ can help the computation
of sets in $\mathcal{C}$? Given any class $\mathcal{C}'$ of helpers,
what class $\mathcal{C} \subseteq \np$ can be helped by sets in
$\mathcal{C}'$? It is worth studying the relationships between
helpers and help-receivers to gain more insights into the notion of
one-sided helping.

A restriction of the notion of one-sided helping, called the concept
of \emph{helping}, was earlier introduced and studied by
Sch{\"{o}}ning~\cite{sch:j:robustness}. A set $A$ is said to
\emph{help} a set $B$ if $B$ is computed by a deterministic oracle
Turing machine $M$ such that on any input $x \in \sigmastar$, (a)
$M^A(x)$ halts in polynomial time, and (b) for all oracles $C$,
$M^{C}(x)$ accepts (though perhaps $M^C(x)$ may take a longer time
than $M^A(x)$ to terminate for $C \neq A$) if and only if $x \in B$.

\begin{definition}[\cite{sch:j:robustness,ko:j:helping}]
\begin{enumerate}
\item A deterministic oracle Turing machine $M$ is \emph{robust} if for
all oracles $A$, $M^A$ halts on each input and $L(M^A) =
L(M^{\emptyset})$.
\item A set $L$ is in the class $\p_{1\hbox{-}\help}(A)$ if there
exists a robust deterministic oracle Turing machine $M$ and a polynomial $p(\cdot)$
such that $L = L(M^{\emptyset})$ and for all $x \in L$, $M^A(x)$ halts in $p(|x|)$ steps.
If $\mathcal{C}$ is a complexity class, then $\p_{1\hbox{-}\help}(\mathcal{C}) =
\bigcup_{A \in \mathcal{C}} \p_{1\hbox{-}\help}(A)$.
\item A set $L$ is in the class $\p_{\help}(A)$ if there
exists a robust deterministic oracle Turing machine $M$ and a polynomial $p(\cdot)$
such that $L = L(M^{\emptyset})$ and for all $x \in \sigmastar$, $M^A(x)$ halts in $p(|x|)$ steps.
If $\mathcal{C}$ is a complexity class, then $\p_{\help}(\mathcal{C}) =
\bigcup_{A \in \mathcal{C}} \p_{\help}(A)$.
\end{enumerate}
\end{definition}

\noindent There has been much investigation on the complexity of
sets that can be one-sided helped by sets belonging to particular
complexity classes. For instance, Ko~\cite{ko:j:helping} proved that
$\np = \p_{1\hbox{-}\help}(\np)$, $\up \subseteq
\p_{1\hbox{-}\help}(\up)$, and $\p_{1\hbox{-}\help}(\bpp) \subseteq
\rp$. Ko~\cite{ko:j:helping} posed the question whether
$\p_{1\hbox{-}\help}(\up)$ is exactly the same as $\up$. Cai,
Hemachandra, and Vysko\v{c}~\cite{cai-hem-vys:b:promise} proved that
relativizable proof techniques cannot resolve this question: There
is a relativized world, where $\p_{1\hbox{-}\help}(\up)$ strictly
contains $\up$~\cite{cai-hem-vys:b:promise}. Cintioli and
Silvestri~\cite{cin-sil:j:helping} strengthened this result of Cai,
Hemachandra, and Vysko\v{c}~\cite{cai-hem-vys:b:promise}. They
exhibited an oracle $A$ such that $\p^A_{\help}(\up^A) \nsubseteq
\few^A$ and an oracle $B$ such that $\p^B_{\help}(\up^B) \nsubseteq
R^{p, B}_{b}(\fewp(n^{\log ^{O(1)}n})^B)$, where $\fewp(n^{\log
^{O(1)}n})$ is the class of all sets accepted by $\nptm$s with at
most $n^{\log ^{O(1)}n}$ accepting paths on inputs of length $n$.
Despite these negative (oracle) results on the provability of
containment of $\p_{\help}(\up)$ in classes such as $\up$, $\fewp$,
and $\few$, Cai, Hemachandra, and
Vysko\v{c}~\cite{cai-hem-vys:b:promise} were successful in obtaining
an exact characterization of $\p_{1\hbox{-}\help}(\up)$. They proved
that $\p_{1\hbox{-}\help}(\up)$ is the closure of $\up$ under
$\leq^{p}_{lpos}$ reductions, where $\leq^{p}_{lpos}$ is the
polynomial-time \emph{locally positive} reduction introduced by
Hemachandra and Jain~\cite{hem-jai:j:pos}.

We generalize and improve the relativized separation of
$\p_{1\hbox{-}\help}(\up)$ from $\up$ by Cai, Hemachandra, and
Vysko\v{c}~\cite{cai-hem-vys:b:promise} in
Corollary~\ref{cor:helping-up-k}. It remains an open question
whether any of the oracle results by Cintioli and
Silvestri~\cite{cin-sil:j:helping}, i.e., existence of oracles $A$
and $B$ such that $\p^A_{\help}(\up^A) \nsubseteq \few^A$ and
$\p^B_{\help}(\up^B) \nsubseteq R^{p, B}_{b}(\fewp(n^{\log
^{O(1)}n})^B)$, imply our result in Corollary~\ref{cor:helping-up-k}
for the case of $h=1$, or vice-versa. It also remains an open
question whether the oracle results by Cintioli and
Silvestri~\cite{cin-sil:j:helping} can be generalized so that they
hold for $\p_{\help}(\up_{\leq h})$, for any $h \geq 1$.

\begin{theorem}
\label{thm:helping-up-k} For all $h \geq 1$, there exists an oracle
$\aor$ such that
\[
R^{p}_{dtt}(\up_{\leq h}^{\aor}) \nsubseteq
R^{p,\aor}_{s,b}(\promiseup^{{{\rm U}}\Sigma_{h-1}^{p,\aor}}).
\]
\end{theorem}
\sproof
The construction of the oracle $\aor$ %
is essentially the same as that in
Theorem~\ref{thm:promis-truth-table-versus-Turing} except with few
minor changes. For each length $n$, we will now (1) reserve a
segment of $n$ regions $S_{n,f} =_{df} 1^n01^f0\Sigma^{4n}$, where
$f \in [n]$, (2) define $S_n =_{df} \bigcup_{f=1}^{n} S_{n,f}$, and
(3) stipulate that $||\aor \cap S_{n,f}|| \leq h$ for all $f \in
[n]$. The rest of the proof is just an imitation of the proof of
Theorem~\ref{thm:promis-truth-table-versus-Turing}.~\qed

\begin{corollary}
\label{cor:helping-up-k} For all integer $h \geq 1$, there exists an
oracle $\aor$ such that
\[
\p_{1\hbox{-}\help}(\up_{\leq h}^{\aor}) \nsubseteq
R^{p,\aor}_{s,b}(\promiseup^{{{\rm U}}\Sigma_{h-1}^{p,\aor}}).
\]
\end{corollary}
\sproof This follows from Theorem~\ref{thm:helping-up-k}, since it
can be easily shown that $R^{p}_{dtt}(\up_{\leq h}) \subseteq
\p_{1\hbox{-}\help}(\up_{\leq h})$ in every relativized
world.\footnote{This same observation was used by Cai, Hemachandra,
and Vysko\v{c}~\cite{cai-hem-vys:b:promise} (for the case of $h=1$)
in constructing an oracle relative to which
$\p_{1\hbox{-}\help}(\up) \nsubseteq \up^{\aor}$.}~\qed

\section{Robust Unambiguity}
\label{sec:robust-unambig-sigmak}

\noindent So far we looked at several applications of
Lemma~\ref{lemma:weakness-uph-machines} in constructing relativized
worlds involving arbitrary levels of the unambiguous polynomial
hierarchy. Lemma~\ref{lemma:weakness-uph-machines}, in essence,
shows the computational limitations of a $\Sigma_{k}(A)$-system
under certain weak restrictions. What if we impose a more stringent
restriction on a $\Sigma_{k}(A)$-system? This question is relevant
to our next investigation.

We study the power of robustly unambiguous $\Sigma_{k}(A)$-system in
Theorem~\ref{thm:robust-unambiguous-sigmak}. (Recall from
Definition~\ref{def:robust-unambig-sigmakA} in
Section~\ref{sec:prelim}, a $\Sigma_{k}(A)$-system $[A; N_1, N_2,
\ldots, N_k]$ is robustly unambiguous if for every oracle $B$,
$[A\oplus B; N_1, N_2, \ldots, N_k]$ is unambiguous.)
Theorem~\ref{thm:robust-unambiguous-sigmak} illustrates the
following fact: A robustly unambiguous $\Sigma_{k}(A)$-system is so
weak that given any oracle set $B$ and input $x$, the hierarchical
nondeterministic polynomial-time oracle access to $B$ in $[A\oplus
B; N_1, N_2, \ldots, N_k](x)$ can be stripped down and turned into a
deterministic polynomial-time oracle access (to $B$) without
changing the decision (i.e, acceptance or rejection) of the
$\Sigma_{k}(A\oplus B)$-system on input $x$. As a corollary, we
obtain a generic oracle collapse of $\uph$ to $\p$ assuming $\p =
\np$. (See,
e.g.~\cite{blu-imp:c:generic,fen-for-kur-li:j:oracle-toolkit} for
generic oracles and concepts related to them.)

\begin{theorem}
\label{thm:robust-unambiguous-sigmak} For all $A \subseteq
\sigmastar$ and $k \geq 1$, if the $\Sigma_k(A)$-system $[A; N_1,
N_2, \ldots, N_k]$ is robustly unambiguous, then for every $B
\subseteq \sigmastar$,
\[
L[A \oplus B; N_1, N_2, \ldots, N_k] \in \p^{\Sigma_{k}^{p,A} \oplus B}.
\]
\end{theorem}

\noindent \sproof The proof is by induction over $k$. Fix an
arbitrary set $A \subseteq \sigmastar$. The case for $k=1$ follows
by relativization of~\cite[Theorem~2.1]{har-hem:j:rob}. So suppose
that $k > 1$. Notice the following facts: For all sets $B \subseteq
\sigmastar$, we have (a) $L[A \oplus B; N_{1}, N_{2}, \ldots, N_{k}]
= L[L[A \oplus B; N_2, N_3, \ldots, N_k]; N_1]$, (b) $[L[A \oplus B;
N_2, N_3, \ldots, N_k]; N_1]$ is unambiguous, and (c) $[A; N_2, N_3,
\ldots, N_k]$ is robustly unambiguous. Thus by induction hypothesis,
we have that for all sets $B$, $L(A,B) =_{df} L[A \oplus B; N_2,
N_3, \ldots, N_k] \in \p^{\Sigma^{p, A}_{k-1} \oplus B}$. We can now
define a nondeterministic polynomial-time Turing machine $N'_1$ and
a set $D \in \Sigma^{p,A}_{k-1}$ such that for all sets $B \subseteq
\sigmastar$,
\begin{eqnarray*}
&& L[L(A,B); N_1] = L[D \oplus B; N'_1], \mbox{ and } [D; N'_1]
\mbox{ is robustly unambiguous}.
\end{eqnarray*}
It follows by the (strong) induction hypothesis that for every set
$B$, $L[D\oplus B; N'_1]$ $\in$ $\p^{\np^{D} \oplus B}$ $\subseteq$
$\p^{\np^{\Sigma^{p, A}_{k-1}} \oplus B}$. Thus the inductive step
is proved, since $L[A \oplus B; N_1, N_2, \ldots, N_k] = L[D \oplus
B; N'_1] \in \p^{\Sigma_{k}^{p,A} \oplus B}$.~\qed

\begin{corollary}
\label{cor:cohen-usigmapk-equal-p} If $\p = \np$, then relative to a
(Cohen) generic $G$, $\p = \uph$.
\end{corollary}

\noindent The last corollary generalizes a result of Blum and
Impagliazzo: If $\p = \np$, then relative to a (Cohen) generic $G$,
$\p^{G} = \up^{G}$~\cite{blu-imp:c:generic}. Fortnow and
Yamakami~\cite{for-yam:j:generic-separation} demonstrated that
similar collapses relative to any (Cohen) generic $G$ do not occur
at higher levels of the polynomial hierarchy. They proved that for
each $k \geq 2$, there exists a tally set in $\up^{\Sigma_{k-1}^{p,
G}, G} \cap \Pi_{k}^{p, G}$ but not in $\p^{\Sigma_{k-1}^{p, G},G}$.
Thus Corollary~\ref{cor:cohen-usigmapk-equal-p} contrasts with this
generic separation by Fortnow and Yamakami.

\section{Conclusion and Open Problems}

\noindent We presented a counting technique to investigate the
structure of relativized hierarchical unambiguous computation.
However, two interesting problems have remained open, for whose
resolutions the technique presented in this paper could be useful.
These problems are:

\begin{enumerate}
\item \textbf{Simultaneous Immunity and Simplicity in the
Relativized Unambiguous Polynomial Hierarchy.}

\noindent Complexity class separations can be evaluated in terms of
their quality. A separation of a complexity class $\mathcal{C}_1$
from another class $\mathcal{C}_2$ by an \emph{immune} set requires
the existence of an infinite set $L$ in $\mathcal{C}_1$ such that no
infinite set in $\mathcal{C}_2$ can be a subset of $L$; the set $L$
is called $\mathcal{C}_2$-\emph{immune} or \emph{immune} to
$\mathcal{C}_2$. A separation of a complexity class $\mathcal{C}_1$
from a class $\mathcal{C}_2$ by a \emph{simple} set requires the
existence of a co-infinite set (i.e., a set whose complement is
infinite) $L$ in $\mathcal{C}_1$ such that $L$ is not in
$\mathcal{C}_2$ and $\overline{L}$ is immune to $\mathcal{C}_1$; the
set $L$ is called $\mathcal{C}_1$-\emph{simple} or \emph{simple} for
$\mathcal{C}_1$. Finally, a separation of a complexity class
$\mathcal{C}_1$ from a class $\mathcal{C}_2$ by a set that is both
\emph{simple and immune} requires the existence of a set $L$ in
$\mathcal{C}_1$ such that $L$ is $\mathcal{C}_1$-simple and
$\mathcal{C}_2$-immune.

\hspace*{1 cm}  An oracle separation of a complexity class from
another class by a set that is both simple and immune is considered
a much more difficult problem than the oracle separations of the
same classes by simple sets or by immune sets alone. This point has
been discussed in~\cite{buh-toren:c:compl-compl}, which we explain
in our own words as follows: ``Intuitively, if a set $L$ is
$\mathcal{C}$-immune, then the set $L$ must have low density since
no infinite set in $\mathcal{C}$ can be a subset of $L$. Similarly,
if a set $L$ is $\mathcal{C}$-simple, then the set $L$ must have
high density since $\overline{L}$ is $\mathcal{C}$-immune.
Consequently, separation of a complexity class $\mathcal{C}_1$ from
another class $\mathcal{C}_2$ by a set $L \in \mathcal{C}_1$ that is
both $\mathcal{C}_1$-simple and $\mathcal{C}_2$-immune requires the
set $L$ to have conflicting requirements: $L$ must be dense enough
so as to be $\mathcal{C}_1$-simple and must be thin enough so as to
be $\mathcal{C}_2$-immune."

\hspace*{1 cm}  Buhrman and
Torenvliet~\cite{buh-toren:c:compl-compl} showed that relative to an
oracle, the first level and the second level of the polynomial
hierarchy separate by sets that are both simple and immune. Using
Kolmogorov complexity for oracle constructions, they proved that
relative to an oracle $A$, $\np$ has a set that is both $\np$-simple
and $(\np \cap \conp)$-immune, and relative to an oracle $B$,
$\Pi_{2}^{p}$ has a set that is both $\Pi_{2}^{p}$-simple and
$(\Sigma^{p}_{2} \cap \Pi^{p}_{2})$-immune. However, it is currently
open whether there is an oracle relative to which the third level or
any higher level of the polynomial hierarchy separates by a set that
is both simple and immune. In fact, it is also open whether there is
an oracle relative to which $\np$ has a set that is both
$\np$-simple and $\conp$-immune.

\hspace*{1 cm} We expect that the situation for the unambiguous
polynomial hierarchy is quite different from the one for the
polynomial hierarchy. We hope that it might be possible that an
application of our proof technique in conjunction with the
Kolmogorov arguments of Buhrman and
Torenvliet~\cite{buh-toren:c:compl-compl} lead to a construction of
an oracle relative to which all the levels of the unambiguous
polynomial hierarchy separate by sets that are both simple and
immune.

\item \textbf{Random Oracle Separation of the Relativized
Unambiguous Polynomial Hierarchy.}

\noindent There has been an abundance of complexity theoretic
results that hold with probability one relative to a random oracle.
Some prominent random oracle results are: (1) probability one
separation of $\np$ from $\p$ with bi-immunity, and of $\np$ from
$\conp$~\cite{ben-gil:j:prob1}, (2) probability one separation of
$\np$ from $\p/{{\rm poly}}$~\cite{lut-sch:j:circuit-prob-one}, (3)
probability one separation of $\conp$ from $\np$ with
immunity~\cite{ver:c:random-oracle}, and (4) probability one
separation of $\pspace$ from $\ph$~\cite{cai:j:prob1,bab:j:prob1}.
Despite so many random oracle results, the probability one
separation of the levels of the polynomial hierarchy relative to a
random oracle is still an open problem. %
(See~\cite{hem-ram-zim:j:worlds-to-die-for} for an extensive
discussion on this problem.) Currently, only the circuit
complexity-theoretic approach is known for separating the higher
levels (levels beyond three) of the polynomial hierarchy, but the
circuit approach has so far not been successful in resolving this
longstanding open problem.

\hspace*{1 cm} We believe that the case of the unambiguous
polynomial hierarchy is easier. In
Theorem~\ref{thm:up-leqk-notin-usigmapk}, we have used our counting
technique to show that for all $k \geq 1$, there is an oracle $A$
such that $\up_{\leq k+1}^{A}$ is not contained in ${{\rm
U}}\Sigma^{p,A}_{k}$. Thus, unlike the case of the polynomial
hierarchy for which only the circuit approach is known for the
relativized separation of all its levels, the levels of the
relativized unambiguous polynomial hierarchy are separable by
counting arguments alone, and thus by completely avoiding the
machinery of circuit complexity. This raises our hope that a
probability one separation of the levels of $\uph$ might be easier
to achieve than its counterpart, i.e., a probability one separation
of the levels of $\ph$ relative to a random oracle.
\end{enumerate}

\bigskip

\noindent \textbf{Acknowledgment} We thank Lane Hemaspaandra and
J\"{o}rg Rothe for their constant encouragement and support.

\bibliography{../../common}

\end{document}